\documentclass[review]{elsarticle}
\journal{Comprehensive Computational Chemistry}

\usepackage{amssymb}
\usepackage{mathtools}
\usepackage{color}
\usepackage{booktabs}
\usepackage{units}
\usepackage{bm}
\usepackage{braket}
\usepackage{diagbox}
\usepackage{caption}
\usepackage{subcaption}
\usepackage[version=3]{mhchem}

\DeclareMathOperator{\Tr}{Tr}

\renewcommand{\vec}[1]{\boldsymbol{#1}}
\newcommand{\mat}[1]{\mathbf{#1}}
\newcommand{\pd}[2]{\frac{\partial #1}{\partial #2}}
\newcommand{\dg}[1]{#1^{\dagger}}

\begin{document}

\begin{frontmatter}

\title{Relativistic Real-Time Methods}


\author[inst1,inst2]{Marius Kadek}
\author[inst1,inst3]{Lukas Konecny}
\author[inst1,inst4]{Michal Repisky\corref{cauthor}}
\cortext[cauthor]{Corresponding Author}
\ead[url]{www.respectprogram.org/michalrepisky}
\ead{michal.repisky@uit.no}

\affiliation[inst1]{organization={Hylleraas Centre for Quantum Molecular Sciences, 
                                  Department of Chemistry, UiT The Arctic University of Norway}, 
                    city={Tromsø},
                    postcode={9037},
                    country={Norway}}

\affiliation[inst2]{organization={Department of Physics, College of Science, Northeastern University}, 
                    city={Boston},
                    postcode={Massachusetts 02115},
                    country={USA}}                    

\affiliation[inst3]{organization={Max Planck Institute for the Structure and Dynamics of Matter, Center for Free Electron Laser Science},
                    city={Hamburg},
                    postcode={22761},
                    country={Germany}}

\affiliation[inst4]{organization={Department of Physical and Theoretical Chemistry, Faculty of Natural Sciences, Comenius University}, 
                    city={Bratislava},
                    postcode={84104},
                    country={Slovakia}}

\begin{abstract}
Recent advances in laser technology enable to follow electronic motion at
its natural time-scale with ultrafast pulses, leading the way towards atto- and 
femtosecond spectroscopic experiments of unprecedented resolution.
Understanding of these laser-driven processes, which almost inevitably involve 
non-linear light–matter interactions and non-equilibrium electron dynamics, 
is challenging and requires a common effort of theory and experiment.
Real-time electronic structure methods provide the most straightforward way to 
simulate experiments and to gain insights into non-equilibrium electronic processes.
In this Chapter, we summarize the fundamental theory underlying the relativistic particle--field interaction Hamiltonian as well as equation-of-motion for exact-state wave function in terms of the one- and two-electron reduced density matrix. Further, we discuss the relativistic real-time electron dynamics mean-field methods with an emphasis on Density-Functional Theory and Gaussian basis, starting from the four-component (Dirac) picture and continue to the two-component (Pauli) picture, where we introduce various flavours of modern exact two-component (X2C) Hamiltonians for real-time electron dynamics. We also overview several numerical techniques for real-time propagation and signal processing in quantum electron dynamics. We close this Chapter by listing selected applications of real-time electron dynamics to frequency-resolved and time-resolved spectroscopies.
\end{abstract}


\end{frontmatter}

\section{Objectives box}

The principal objectives of this Chapter are:
\begin{itemize}
    \item Introduction to the fundamental theory leading to the relativistic particle--field interaction Hamiltonian.
    \item Discussion of the equations-of-motion for exact-state wave function in terms of the one-electron and two-electron reduced density matrix.
    \item Introduction to the relativistic four-component real-time electron dynamics mean-field methods with an emphasis on Density-Functional Theory and Gaussian basis.
    \item Detailed overview of various exact two-component (X2C) transformations towards the relativistic two-component real-time electron dynamics.
    \item Overview of numerical techniques for real-time propagation and signal processing in quantum electron dynamics. 
    \item Selected application of real-time electron dynamics to frequency-resolved and time-resolved spectroscopies.
\end{itemize}

\section{Introduction}

The rapid advancement of laser technology in the past decades allows us to probe matter on
spatiotemporal scales that approach the characteristic time and length scales of the electron,
opening the field of attosecond science~\cite{Corkum2007,Nisoli2017}. This development has forced quantum chemists to shift their attention from the time-independent to the time-dependent Schrödinger or Dirac equation. Real-time electronic structure theory thus describes the explicit time-evolution of the wave function or the electron density driven by non-equilibrium condition of the Hamiltonian under external perturbation(s). All physical quantities of a molecular system are then extracted non-perturbatively from the time-varying part of the wave function or the electron density. Due to non-perturbative nature, real-time methods represent the most straightforward approach to dynamical property calculations and enable the use of external perturbations of arbitrary strength, shape and duration, capturing in general both linear and non-linear effects within a wide spectral window from a single run~\cite{Nisoli2017,li2020real,SverdrupOfstad2023}. This distinguishes real-time electronic structure theory from response theory where all physical quantities are obtained in the frequency domain using perturbation expansion~\cite{helgaker2012recent}.

Historically, an early work on explicit time-propagation of electronic wave function dates back to 1990 when Cederbaum and coworkers developed the nonrelativistic multiconfigurational time-dependent Hartree (MCTDH) method~\cite{Meyer1990}. Shortly after, Micha and Runge developed a real-time time-dependent Hartree--Fock (RT-TDHF) approach that couples electronic and nuclear motions~\cite{Micha1994}, whereas Theilhaber~\cite{theilhaber1992}, and Yabana and Bertsch~\cite{yabana1996}, introduced the first-ever real-time time-dependent density functional theory (RT-TDDFT) combining the local density approximation with real-space grid methodology. In the condensed matter physics community, these pioneering works led to several implementations of the time-propagation formalism using either localized basis sets~\cite{Tsolakidis2002,hekele2021all}, plane waves~\cite{Qian2006,Walker2007,Walter2008}, or real-space grids~\cite{castro2006}. Advancements in computing power and numerical algorithms have enabled performing large-scale RT-TDDFT simulations even on periodic solids~\cite{Schleife2012,Andrade2012,Andermatt2016,Shepard2021,pemmaraju2020simulation}. In the quantum chemistry community, the first nonrelativistic RT-TDDFT implementation based on popular Gaussian-type atomic orbitals was pioneered by Isborn and coworkers in 2007~\cite{Isborn2007} and later adopted by several other groups~\cite{lopata2011modeling,Gao2012}. The extension of RT-TDDFT to the relativistic four-component (4c) realm was presented by Repisky and coworkers in 2015~\cite{repisky2015excitation,Kadek2015}. As shown by these authors, significant gains in computer time was obtained by transforming the parent 4c RT-TDDFT to an exact two-component (X2C) form~\cite{Konecny2016}, although the accuracy of reference 4c results was achieved only after inclusion of the two-electron and exchange–correlation picture-change corrections~\cite{moitra2023accurate,Konecny2023}. Beyond DFT, there has been growing interest in explicit time propagation of correlated methods such as multiconfigurational self-consistent-field~\cite{Sato2013,Miyagi2013,Miyagi2014,Sato2015}, configuration interaction~\cite{CI1,CI2,CI3,CI4,CI5,CI6,CI7}, algebraic diagrammatic construction~\cite{ADC1,ADC2,ADC3,ADC4,ADC5,ADC6,ADC7}, density matrix renormalization group\cite{Baiardi2021}, Møller-Plesset~\cite{Kristiansen2022}, and coupled cluster~\cite{SverdrupOfstad2023,CC0,CC1,CC2,CC3,CC4,CC5,CC6,CC7,CC8,CC9,CC10,CC11,CC12} theories.

At the non-relativistic level of theory, a plethora of applications of real-time methods has been presented including 
UV/Vis absorption spectroscopy~\cite{yabana1996, Yabana2006, lopata2011modeling},
excited-state absorption~\cite{Fischer2015excited},
photoionization~\cite{ullrich1998electron, deWijn2008strong},
X-ray absorption~\cite{Akama2010, Lopata2012},
chiroptical spectroscopies~\cite{varsano2009towards, pipolo2014cavity, mattiat2019electronic, mattiat2022comparison, monti2023electronic},
non-linear optical properties~\cite{takimoto2007real, Ding2013},
spin and magnetization dynamics~\cite{Ding2014AbInitio, peralta2015magnetization},
molecular conductance~\cite{cheng2006simulating},
pump-probe spectroscopy~\cite{DeGiovannini2013},
photoinduced electric currents~\cite{Nobusada2007},
plasmon resonances~\cite{Gao2012},
singlet--triplet transitions~\cite{isborn2009singlet},
and magnetic circular dichroism~\cite{Lee2011}.
This list of applications is by no means exhaustive and a reader interested in a more
thorough exploration of the use of non-relativistic real-time methods is referred
to the recent review~\cite{li2020real} and references therein.

The advent of soft X-ray free electron laser pulses with subfemtosecond temporal widths have opened new ways to investigate time-resolved dynamics involving inner-shell electrons. A prerequisite for reliable quantum-chemical modeling of these processes is the inclusion of relativistic effects, defined as differences between the exact Dirac (four-component) description of matter and an approximate Schrödinger (one-component) description. This requirement stems from the fact that the inner-shell orbitals involved in X-ray absorption/emission processes are most affected by relativity, manifestations of which are frequency shifts of spectral lines due to the scalar (SC) relativistic effects as well as spectral fine structure splitting arising from the spin-orbit (SO) coupling~\cite{Kadek2015,Konecny2023}. The relativistic effects are significant even in light (third row) elements~\cite{Kadek2015} and increase in importance for heavier elements~\cite{Konecny2023}, highlighting the need for relativistic description across the Periodic Table. Therefore, the most accurate way to perform real-time simulations is the use of full four-component (4c) Dirac formalism where both the SC and SO relativistic effects are included variationally. The first 4c extension of the time-propagation formalism was presented by Repisky and coworkers at the RT-TDDFT level~\cite{repisky2015excitation,Kadek2015}, involving the program package ReSpect~\cite{ReSpect}. Recently, De Santis and coworkers reported a similar 4c RT-TDDFT implementation in the BERTHA code~\cite{DeSantis2020pyberthart}. While advancements in computing power and numerical algorithms have enabled performing fairly large 4c real-time electron dynamics simulations~\cite{repisky2015excitation,Kadek2015,Konecny2016,moitra2023accurate,Konecny-JCP-149-204104-2018,DeSantis2020pyberthart}, there is still interest in developing approximate two-component (2c) methods that maintain the accuracy of the parent 4c method at a fraction of its computational cost. In this respect, the X2C Hamiltonian has gained wide popularity in quantum chemistry community as it reduces the original 4c problem by half at the expense of only a few simple algebraic manipulations~\cite{Heully1986,Jensen2005,Kutzelnigg2005,Liu2007,Ilias2007}. As shown independently by Konecny~\cite{Konecny2016} and Goings~\cite{Goings2016}, the central idea of X2C transformation can be extended to the real-time electron dynamics framework, provided the X2C decoupling matrix satisfies an adiabatic approximation~\cite{Konecny2016}.
However, both real-time X2C implementations utilize a crude one-electron X2C (1eX2C) Hamiltonian model, which typically leads to absolute errors for core spinor energies of heavier elements of the order of tens of Hartree~\cite{Knecht2022}. As shown by Knecht and coworkers~\cite{Knecht2022}, accuracy of X2C Hamiltonians severely depends on the two-electron and exchange-correlation picture-change correction models employed, and can vary as much as 5-6 orders of magnitude for core-shell energies. As a remedy, the authors introduced two simple yet computationally efficient and numerically accurate X2C Hamiltonian models, dubbed as amfX2C and e(xtended)amfX2C, to correct both SC and SO two-electron and exchange-correlation picture-change effects using simple atomic mean-field quantities, achieving a consistent $\approx10^{-5}$ Hartree/atom accuracy~\cite{Knecht2022}. The theoretical extension and numerical assessment of (e)amfX2C Hamiltonian models was recently performed for conventional and time-resolved TDDFT by Repisky and coworkers~\cite{moitra2023accurate,Konecny2023}. In addition to the previous works, this Chapter also provides an in-depth discussion on the transformation of the original 4c equation-of-motion to its 2c form, particularly focusing on the modern exact two-component (X2C) formalism.

At the relativistic level, the real-time applications are scarcer due to fewer computer programs providing such functionality. These programs include ReSpect~\cite{repisky2015excitation,ReSpect},
Gaussian~\cite{Goings2016},
Chronus Quantum~\cite{williams2020chronus},
PyBerthaRT~\cite{DeSantis2020pyberthart},
BDF~\cite{Ye2022self}, 
and FHI-aims~\cite{hekele2021all}.
The molecular properties addressed at the relativistic level include
UV/Vis absorption spectroscopy~\cite{repisky2015excitation, Goings2016},
X-ray absorption~\cite{Kadek2015, Kasper2018Modeling, Ye2022self},
non-linear optical properties~\cite{Konecny2016},
chiroptical spectroscopies~\cite{Konecny-JCP-149-204104-2018},
high harmonic generation~\cite{DeSantis2020pyberthart},
and pump-probe spectroscopy~\cite{moitra2023accurate}.

Before closing this Section, let us emphasize that we restrict ourselves to (i) the Born--Oppenheimer approximation and therefore coupled electron--nuclear dynamics is not considered here; (ii) the semiclassical approximation where electronic degrees of freedom are described quantum mechanically, while electromagnetic fields are treated classically. In next Sections, we introduce the fundamental theory behind the relativistic particle--field interaction Hamiltonian, and discuss the equation-of-motion for exact-state wave function in terms of the one-electron and two-electron reduced density matrix. Later, we dive into the relativistic four-component real-time electron dynamics mean-field methods with an emphasis on Density-Functional Theory and Gaussian basis, followed by a detailed overview of various exact two-component (X2C) transformation models within the time domain. Finally, we offer a brief overview of numerical techniques for real-time propagation and signal processing, and close this Chapter by listing selected applications in relativistic quantum electron dynamics.

\section{Relativistic Particle--Field Interaction Hamiltonians}
\label{sec:relatHamiltonian}
For the theoretical description of spectroscopic processes, quantum chemistry commonly employs a semiclassical theory. In this framework, the molecules are described quantum-mechanically, whereas the electromagnetic (photon) field is treated classically. This assumption is justified in the limit of large photon numbers -- to be specific, when the photon density exceeds one per cubic wavelength (for a discussion, see Ref.~[\citenum{Sakurai1967}]). If this is not the case, one may need to quantize the photon field as well and work within the framework of quantum electrodynamics~\cite{Craig1988}. To provide an illustrative example, let us consider a laser pulse with intensity $10^{14}\unit{W/cm^{2}}$ and wavelength $\lambda=1064\unit{nm}$. The number of photons per cubic wavelength is then given by ($\hbar$ and $c$ denote the reduced Planck constant and the speed of light, respectively):
\begin{equation}
    \frac{\text{energy flux}}{\hbar\omega}\frac{V}{c}
    =
    \frac{10^{14}\unit{W/cm^{2}}}{1.86\times10^{-19}\unit{J}}
    \frac{1.064^{3}\times10^{-12}\unit{cm^{3}}}{3\times10^{10}\unit{cm/s}}
    \approx
    2\times10^{10},
\end{equation}
which is obviously much greater than one. Therefore, this semiclassical theoretical framework is appropriate for absorption and emission processes, and we rely on this framework throughout this Chapter.

Before we actually dive into particle--field interactions, let us first consider a $N$-electron system alone, \emph{i.e.} in the absence of any electromagnetic (photon) field. In this case, the system is governed by the relativistic electronic Hamiltonian
\begin{equation}
   \label{eq:H}
   \hat{H}
   = 
   \sum_{i}^{N} \hat{h}^\mathrm{D}_{i}
   +
   \frac{1}{2} \sum_{i\neq j}^{N} \hat{g}_{ij}
   .
\end{equation}
Here, $\hat{h}^\mathrm{D}_{i}$ is the famous relativistic Dirac Hamiltonian of a single electron $i$, while $\hat{g}_{ij}$ is the interaction Hamiltonian between electrons $i$ and $j$. A factor one half in front of $\hat{g}$ corrects for double counting of the two-electron interactions. $\hat{h}^\mathrm{D}$ describes the relativistic kinetic energy of an electron as well as its interaction energy with the electrostatic scalar potential $\phi_{0}(\vec{r})$ due to the fixed atomic nuclei. It bears the $4\!\times\!4$ matrix form~\cite{Dirac1928a,Dirac1928b}
\begin{equation} 
  \label{eq:hD}
  \begin{split}
     \hat{h}^\mathrm{D}_{i}
     & = 
     \beta'_{i}m_{e}c^{2} 
     + 
     c\big(\vec{\alpha}_{i}\cdot\vec{p}_{i}\big) 
     - 
     e\phi_{0}(\vec{r}_{i})\mathbb{I}_{4}
    .
  \end{split}
\end{equation}
Here, $\vec{r}_{i}$ and $\vec{p}_{i}\!=\!-i\hbar\vec{\nabla}_{i}$ refer to the position and canonical momentum of the $i$th electron, respectively. $\mathbb{I}_{4}$ is a $4\!\times\!4$ identity matrix, and $-e$, $m_{e}$ and $c$ are constants referring to the electron charge, electron mass and the speed of light in vacuum. When compared to the original expression of Dirac~\cite{Dirac1928a,Dirac1928b}, $\hat{h}^\mathrm{D}$ utilizes the reduced rest mass energy $\beta'm_{e}c^{2}$ with $\beta'\equiv\beta-\mathbb{I}_{4}$ to align the relativistic and non-relativistic energy scales. $\beta$ is one of four new $4\!\times\!4$ matrix variables  
\begin{align}
   \label{eq:alpha+beta}
   \beta =
   \begin{pmatrix}
     \mathbb{I}_{2} &  0_{2} \\
     0_{2}          & -\mathbb{I}_{2}
   \end{pmatrix};
   \qquad
   \vec\alpha =
   \begin{pmatrix}
     0_{2} & \vec{\sigma}\\
     \vec{\sigma} & 0_{2}
   \end{pmatrix},
\end{align}
introduced by Dirac to formulate relativistic quantum-mechanical equations of motion for spin-1/2 particles that are linear in space and time~\cite{Dirac1928a,Dirac1928b}. These variables fulfill the anti-commutation relations 
\begin{equation}
   [\alpha_{k},\beta]_{+} = 0_{4};
   \qquad
   [\alpha_{k},\alpha_{l}]_{+} = 2\delta_{kl}\mathbb{I}_{4};
   \qquad
   k,l\in x,y,z,
\end{equation}
and are customarily written in terms of the two-component Pauli spin matrices
\begin{align}
   \label{eq:sigmas}
   \sigma_{x} =
   \begin{pmatrix}
     0 & 1 \\
     1 & 0
   \end{pmatrix};
   \quad
   \sigma_{y} =
   \begin{pmatrix}
     0 & -\mathrm{i}\\
     \mathrm{i} & 0
   \end{pmatrix};
   \quad
   \sigma_{z} =
   \begin{pmatrix}
     1 & 0 \\
     0 & -1
   \end{pmatrix}.
\end{align}
For further reading on the properties and physical interpretation of the Dirac Hamiltonian, the reader is referred to several excellent quantum chemistry textbooks~\cite{Moss1973, Dyall2007, Reiher2014}.

Now, let us subject the $N$-electron system to a classical electromagnetic radiation characterized by the fundamental electromagnetic field vectors: the electric field $\vec{E}\equiv\vec{E}(\vec{r},t)$ and the magnetic field $\vec{B}\equiv\vec{B}(\vec{r},t)$. These vectors satisfy the microscopic Maxwell's equations, which are the basic equations of motion of electromagnetism where charged particles appear as sources. Here, we apply the perturbation theory viewpoint of quantum electrodynamics: to first order it is assumed that the particles of whose motion is being studied do not affect the radiation field, which thus appear as a "driving field"~\cite{Craig1988}. Therefore, assuming that the sources of the radiation field are sufficiently remote from a molecule of interest, the $\vec{E}$ and $\vec{B}$ fields are source- and divergence-free, and conveniently described in terms of the scalar potential $\phi\equiv\phi(\vec{r},t)$ and the vector potential $\vec{A}\equiv\vec{A}(\vec{r},t)$, satisfying~\cite{Jackson1998}:
\begin{equation}
    \label{eq:E+B}
    \begin{split}
       \vec{E}(\vec{r},t) &= -\vec{\nabla}\phi(\vec{r},t) - \frac{\partial \vec{A}(\vec{r},t)}{\partial t}
       ;\qquad
       \vec{\nabla}\cdot\vec{E}(\vec{r},t) = 0;
       \\
       \vec{B}(\vec{r},t) &= \vec{\nabla}\times\vec{A}(\vec{r},t)
       ;\qquad\qquad\qquad~
       \vec{\nabla}\cdot\vec{B}(\vec{r},t) = 0.
    \end{split}
\end{equation}

In fact, both electromagnetic potentials enter the Dirac Hamiltonian and describe the coupling of an electron to the classical electromagnetic field as~\cite{Dirac1928b}
\begin{equation} 
  \label{eq:hD+hint}
  \begin{split}
     \hat{h}^\mathrm{D}_{i}(t)
     & = 
     \beta'_{i}m_{e}c^{2} 
     + 
     c\big(\vec{\alpha}_{i}\cdot\vec{p}_{i}\big)
     -
     e\phi_{0}(\vec{r}_{i})\mathbb{I}_{4}
     - 
     e\phi(\vec{r}_{i},t)\mathbb{I}_{4}
     + 
     ec\big(\vec{\alpha}_{i}\cdot\vec{A}(\vec{r}_{i},t)\big)
    .
  \end{split}
\end{equation}
When compared to the field-free Dirac Hamiltonian in Eq.~\eqref{eq:hD}, the scalar electrostatic potential due to the nuclei $\phi_{0}(\vec{r})$ is substituted by the time-dependent potential: $\phi_{0}(\vec{r}) \rightarrow \phi_{0}(\vec{r}) + \phi(\vec{r},t)$, and the canonical momentum of an electron $\vec{p}$ is substituted by the mechanical momentum: $\vec{p} \rightarrow \vec{p}+e\vec{A}(\vec{r},t)$. The latter substitution is known in literature as the principle of minimal electromagnetic coupling substitution~\cite{Gell1956}. 

To gain insights into the physical interpretation of the matter--field interaction, let us consider the expectation value of the relativistic one-electron interaction Hamiltonian given by last two terms in Eq.~\eqref{eq:hD+hint}
\begin{equation} 
  \label{eq:hint}
  \begin{split}
     \int
     \psi^{\dagger}(\vec{r},t)
     \Big[
     - 
     e\phi(\vec{r},t)\mathbb{I}_{4}
     + 
     ec\big(\vec{\alpha}\cdot\vec{A}(\vec{r},t)\big)
     \Big]
     \psi(\vec{r},t)
     d^{3}\vec{r}
     =
     \\
     =
     \int
     \Big[
     \rho(\vec{r},t)\phi(\vec{r},t)
     -
     \vec{j}(\vec{r},t)\cdot\vec{A}(\vec{r},t)
     \Big]
     d^{3}\vec{r}
    .
  \end{split}
\end{equation}
 Assuming multiplicative potentials $\phi$ and $\vec{A}$, the second equation reveals that the scalar potential is coupled to the electron charge density $\rho$ -- \emph{i.e.}, the charge of the electron times its probability distribution
 \begin{equation} 
   \label{eq:defDensity01e}
   \rho(\vec{r},t)
   =
   -e\psi^{\dagger}(\vec{r},t)
   \mathbb{I}_{4}
   \psi(\vec{r},t)
   ,
\end{equation}
whereas the vector potential is coupled to the electron current density $\vec{j}$ -- \emph{i.e.} the charge of the electron times its velocity distribution
\begin{equation} 
   \label{eq:defDensityCurrent1e}
   \vec{j}(\vec{r},t)
   =
   -e\psi^{\dagger}(\vec{r},t)
   c\vec{\alpha}
   \psi(\vec{r},t)
   .
\end{equation}

In order to write the interaction Hamiltonian in its explicit form, we need to know analytical expressions for both potentials $\phi$ and $\vec{A}$. By the use of Maxwell's equations for the source-free field, it can be shown that these potentials satisfy~\cite{Jackson1998}
\begin{equation}
\label{eq:WE-general}
\begin{split}
    \nabla^{2}\phi
    + 
    \frac{\partial}{\partial t}(\vec{\nabla}\cdot\vec{A}) 
    = 0,
    \\
    \nabla^{2}\vec{A} - \frac{1}{c^{2}}\frac{\partial^{2}\vec{A}}{\partial t^{2}} 
    -
    \vec{\nabla} \Big( \vec{\nabla}\cdot\vec{A} + \frac{1}{c^{2}}\frac{\partial\phi}{\partial t} \Big)
    = 0.
\end{split}
\end{equation}
However, there exists a certain arbitrariness in the definition of the potentials, in that it is possible to shift them by the transformation
\begin{equation}
    \label{eq:gauge-transformation}
    \phi \rightarrow \phi' = \phi - \frac{\partial\chi}{\partial t}
    ;\qquad
    \vec{A} \rightarrow \vec{A}' = \vec{A} + \vec{\nabla}\chi
    ,
\end{equation}
where $\chi\equiv\chi(\vec{r},t)$ is an arbitrary scalar function of space and time coordinates called a gauge function. Since the physics, \emph{i.e.} the force law and Maxwell's equations, is sensitive only to the electric field $\vec{E}$ and the magnetic field $\vec{B}$, the transformation of potentials, called a gauge transformation, does not affect it. This is known in physics as gauge invariance and may be readily verified by inserting two pairs of potentials $(\phi,\vec{A})$ and $(\phi',\vec{A}')$ into the expression in Eq.~\eqref{eq:E+B}. In addition, gauge invariance may be exploited to simplify Eq.~\eqref{eq:WE-general}, and this means also the interaction Hamiltonian.

In quantum chemistry, the gauge freedom is fixed by choosing the so-called Coulomb gauge defined by the condition~\cite{Gell1956}
\begin{equation}
   \label{eq:CG}
   \vec{\nabla}\cdot\vec{A}(\vec{r},t) = 0.
\end{equation}
With this condition and the fact that the electric field is divergence-free in free space~\eqref{eq:E+B}, the scalar potential is a constant, \emph{i.e.} $\phi(\vec{r},t)=\phi$, and may be taken as zero to satisfy $|\phi|\rightarrow0$ at spatial infinity. In this case, the equations of motion for electromagnetic potentials~\eqref{eq:WE-general} simplify to
\begin{equation}
    \label{eq:WE-CG}
    \begin{split}
    \nabla^{2}\phi = \phi = 0,
    \\
    \Big( \nabla^{2} - \frac{1}{c^{2}}\frac{\partial^{2}}{\partial t^{2}} \Big) \vec{A}(\vec{r},t) = 0.        
    \end{split}
\end{equation}
The wave equation for the vector potential is identical in form in many problems of wave motion, with a real solution in the form of a monochromatic, linearly polarized electromagnetic plane-wave~\cite{Jackson1998} 
\begin{equation}
   \label{eq:PW}
   \vec{A}(\vec{r},t)
   =
   \vec{A}_{0}\cos(\vec{k}\cdot\vec{r} - \omega t)
   ,
\end{equation}
where $\vec{A}_{0}$ is a constant real vector called amplitude factor. The argument of the cosine function is called the \emph{phase} of $\vec{A}$ and is given in terms of the wave vector $\vec{k}$ (characterizing the direction of wave propagation) and the angular frequency $\omega$. Note that the phase sometimes contains a phase constant, manipulation of which the cosine function can be converted to a sine function. By substitution of the solution Eq.~\eqref{eq:PW} back into the wave equation Eq.~\eqref{eq:WE-CG}, we find that the magnitude of $\vec{k}$ obeys $|\vec{k}| = k = \omega/c$. In addition, noting that the angular frequency is $\omega=2\pi\nu=2\pi c/\lambda$ with the frequency $\nu$ and wavelength $\lambda$, we also find $|\vec{k}| = k = 2\pi/\lambda$.

To summarize, time evolution of a $N$-electron system subjected to a classical electromagnetic field is governed by the electronic Hamiltonian 
\begin{equation}
   \label{eq:Ht}
   \hat{H}(t)
   = 
   \sum_{i}^{N} \hat{h}^\mathrm{D}_{i}(t)
   +
   \frac{1}{2} \sum_{i\neq j}^{N} \hat{g}_{ij}
   ,
\end{equation}
where its one-electron part given in the Dirac's relativistic formalism as
\begin{equation} 
  \label{eq:hD+hint-CG}
  \begin{split}
     \hat{h}^\mathrm{D}_{i}(t)
     & = 
     \beta'_{i}m_{e}c^{2} 
     + 
     c\big(\vec{\alpha}_{i}\cdot\vec{p}_{i}\big)
     -
     e\phi_{0}(\vec{r}_{i})\mathbb{I}_{4}
     +
     \hat{h}^\mathrm{(v)}(\vec{r}_{i},t)
     ,
  \end{split}
\end{equation}
contains also the electron--field interaction Hamiltonian characterized in the Coulomb gauge entirely by the vector potential
\begin{equation}
  \label{eq:hint-velocity}
  \begin{split}
     \hat{h}^\mathrm{(v)}(\vec{r}_{i},t)
     & =
     ec\vec{\alpha}_{i}\cdot\vec{A}(\vec{r}_{i},t)
     =
     ec\big(\vec{\alpha}_{i}\cdot\vec{A}_{0})\cos(\vec{k}\cdot\vec{r}_{i} - \omega t)
     .
  \end{split}
\end{equation}
In the literature, $\hat{h}^\mathrm{(v)}$ is known as the one-electron interaction Hamiltonian in \emph{velocity representation} which we shall label with the superscript (v). 

Note that the spatial phase of $\hat{h}^\mathrm{(v)}$ can be simplified by considering that wavelengths of electromagnetic waves in the ultraviolet or visible range are very large compared with the spatial extent of typical molecular systems under study. To provide an illustrative example, let us consider a laser pulse with wavelength $\lambda=1064\unit{nm}$ applied to a molecule of size $|\vec{r}| = r = 10\unit{\AA}$. Hence, 
\begin{equation}
   \vec{k}\cdot\vec{r} \leq kr = \frac{2\pi}{1064\unit{nm}}1\unit{nm} \ll 1,
\end{equation}
which implies that the spatial phase of an oscillating electromagnetic wave can be approximated by a constant over the length scale of a molecule (or more precisely over the mean-value of an electronic position), \emph{i.e.}
\begin{equation}
   \exp{[i(\vec{k}\cdot\vec{r})]} 
   =
   1 + i(\vec{k}\cdot\vec{r}) - \frac{1}{2}(\vec{k}\cdot\vec{r})^{2} + ...
   \approx 1
   .
\end{equation}
Independence of the electromagnetic wave on a spatial coordinate is known as \emph{dipole} (or \emph{long-wavelength}) approximation, which brings the velocity interaction Hamiltonian in Eq.~\eqref{eq:hint-velocity} into a particularly simple form labelled here as (vd)
\begin{equation}
  \label{eq:hint-velocity+dipole}
  \begin{split}
     \hat{h}^\mathrm{(v)}(\vec{r}_{i},t)
     \approx
     \hat{h}^\mathrm{(vd)}_{i}(t)
     =
     ec\vec{\alpha}_{i}\cdot\vec{A}(t)
     =
     \frac{ec}{2}\big(\vec{\alpha}_{i}\cdot\vec{A}_{0}\big)
     \big[\exp(-i\omega t) + \exp(i\omega t)\big]
     .
  \end{split}
\end{equation}
Here, we used $\cos(x)=[\exp(ix)+\exp(-ix)]/2$. 
However, special care has to be taken for short wavelengths used for instance in hard X-ray spectroscopy where the dipole approximation may not be adequate. In particular, this is true for heavy-element K-edge X-ray absorption spectroscopy~\cite{Bernadotte2012,List2016}. By including higher-order powers of $\vec{k}\cdot\vec{r}$ in the expansion, one gets multipolar contributions known as electric-quadrupole, magnetic-dipole, etc., and there exist techniques to include these contributions into quantum-chemical calculations~\cite{Bernadotte2012,List2016,List2020,Foglia2022}.  

Before we close this section, let us mention that there exists a unitary transformation of the wave function which yields an altered form of the interaction Hamiltonian that may be more useful for practical calculations. Let us start from the time-dependent Schr\"odinger/Dirac equation with the electronic Hamiltonian $\hat{H}$ containing the velocity-dipole interaction Hamiltonian ($\hat{h}^\mathrm{(vd)}$) given by Eq.~\eqref{eq:hint-velocity+dipole}:
\begin{equation}
   \left( i\hbar\frac{\partial}{\partial t} - \hat{H} \right) \Psi = 0
   ;\qquad
   \hat{H} \equiv \hat{H}(t)
   = 
   \sum_{i}^{N} \big[ \hat{h}^\mathrm{D}_{i} + \hat{h}^\mathrm{(vd)}_{i}(t) \big]
   +
   \frac{1}{2} \sum_{i\neq j}^{N} \hat{g}_{ij}
   .
\end{equation}
 The wave function $\Psi\!\equiv\!\Psi(t)$ can undergo a unitary (gauge) transformation with a freely chosen function $\Lambda\equiv\Lambda(t)$
\begin{equation}
    \Psi = \exp(-i\Lambda)\Psi'. 
\end{equation}
The new wave function $\Psi'\!\equiv\!\Psi'(t)$ is as physically meaningful as the old one, provided
\begin{equation}
   \left( i\hbar\frac{\partial}{\partial t} - \hat{H}' \right) \Psi' = 0
   ;\qquad
   \hat{H}' \equiv \hat{H}'(t)
   = 
   \exp(i\Lambda)\hat{H}(t)\exp(-i\Lambda) - \hbar\frac{\partial\Lambda}{\partial t}
   .
\end{equation}
Now, by selecting $\Lambda$ as
\begin{equation}
    \Lambda(t) 
    = 
    \frac{e}{\hbar}\sum_{i}^{N} \mathbb{I}_{4}\vec{r}_{i}\cdot\vec{A}(t)
    ,
\end{equation}
one replaces the velocity-dipole interaction Hamiltonian in the original electronic Hamiltonian $\hat{H}$ by a new interaction Hamiltonian $\hat{h}^\mathrm{(ld)}$ in the so-called \emph{length-dipole} representation (ld) in the new electronic Hamiltonian $\hat{H}'$:
\begin{equation}
   \label{eq:hint-length+dipole}
   H'(t)
   = 
   \sum_{i}^{N} \big[ \hat{h}^\mathrm{D}_{i} + \hat{h}^\mathrm{(ld)}_{i}(t) \big]
   +
   \frac{1}{2} \sum_{i\neq j}^{N} \hat{g}_{ij}
   ;\qquad
   \hat{h}^\mathrm{(ld)}_{i}(t)
   =
   e\mathbb{I}_{4}\vec{r}_{i}\cdot\vec{E}(t)
   .
\end{equation}
Physically, $\hat{h}^\mathrm{(ld)}$ couples the classical electric field $\vec{E}(t)$ defined as
\begin{equation}
   \vec{E}(t) = - \frac{\partial}{\partial t} \vec{A}(t)
   ,
\end{equation}
to the quantum-mechanical system characterized by the sum of the electric dipole moment operators of individual electrons ($\vec{\mu}_{i}=-e\vec{r}_{i}\mathbb{I}_{4}$). This gauge transformation was first discussed at the nonrelativistic level by G\"oppert-Mayer in 1931~\cite{GoppertMayer1931} and therefore it is often named after her. An additional reading on quantum-mechanical gauge invariance and general unitary transformations for atoms and molecules in interactions with radiation can be found in Ref.~[\citenum{Bandrauk2013}].

\section{Equations-of-motion for exact-state wave function}

\subsection{Time-dependent Schr\"odinger equation}

In the most general case, the time evolution of a quantum-mechanical system is governed by the time-dependent equation-of-motion
\begin{equation}
    i\hbar\pd{\Psi(t)}{t} = \hat{H}(t)\Psi(t),
\label{eq:eomexact}
\end{equation}
where $\hat{H}(t)$ is the Hamiltonian operator that is explicitly dependent on time via external electromagnetic fields. If we are interested in the response of molecules subjected to ultrafast laser pulses and similar processes occurring at atto- or femtosecond time scales, we can study electron dynamics decoupled from the nuclear motion. However, molecular vibrations and nuclear relaxation occur on a time scale of 10--100 fs, and in principle should be included in the computation. Nevertheless, performing electron dynamics simulations with fixed nuclear configuration at this time scale is still beneficial for improving the spectral resolution and aids analyzing electron excitations without the effect of nuclear dynamics. In such a case, $\Psi(t)$ in Eq.~\eqref{eq:eomexact} is the many-electron wave function depending on the position $\vec{r}_i$ and spin of all electrons, and the Hamiltonian $\hat{H}$ is the many-electron Hamiltonian defined in Eq.~\eqref{eq:Ht} containing the electron kinetic operator, Coulomb interactions between electrons and nuclei with fixed positions, and interactions between the system and external electromagnetic fields discussed in more detail in Section~\ref{sec:relatHamiltonian}.

Eq.~\eqref{eq:eomexact} also remains valid in relativistic case, provided additional approximations are assumed. The Hamiltonain $\hat{H}$ needs to be treated as a multicomponent operator acting on multicomponent wave functions to reflect the fact that in relativistic theory, electron spin and orbital degrees of freedom interact with each other via the spin--orbit coupling terms. However, in a truly relativistic picture, we would need to consider multiple time variables associated with each electron's frame of reference. Such effects arising from the relative time are always neglected when studying molecular systems, and Eq.~\eqref{eq:eomexact} thus assumes the absolute time approximation, which leads to a single time variable $t$. For further discussion on the relativistic theory of many electrons, see Ref.~\cite{Reiher2014}.

Response of the system to external time-dependent electromagnetic fields can be studied by solving Eq.~\eqref{eq:eomexact}. This can be achieved by using the formalism of response theory~\cite{thorvaldsen2008density,helgaker2012recent}, in case the external fields are weak and can be regarded as small perturbations to the system compared to the intrinsic unperturbed Hamiltonian. Alternatively, the equation can be solved numerically by propagating the wave function in real time, which facilitates studying processes that involve arbitrarily strong fields.

\subsection{Reduced density matrices}
\label{sec:RDMs}

Since the many-electron wave function is a complicated object that depends on the spatial coordinates of each electron, for the forthcoming discussion, it will be more convenient to work in the formalism of reduced density matrices (RDMs)~\cite{mcweeny1989method}. In the time domain, we can define the one-electron and two-electron RDMs, respectively, as
\begin{equation}
    D(\vec{r}_1;\vec{r}'_1;t) = N \int \Psi(\vec{r}_1,x_2,\ldots,x_N,t) \dg{\Psi}(\vec{r}'_1,x_2,\ldots,x_N,t) dx_2\ldots dx_N,
\label{eq:def1RDM}
\end{equation}
and
\begin{multline}
    \Gamma(\vec{r}_1,\vec{r}_2;\vec{r}'_1,\vec{r}'_2;t) = N(N-1) \int \Psi(\vec{r}_1,\vec{r}_2,x_3,\ldots,x_N,t) \\
    \times \dg{\Psi}(\vec{r}'_1,\vec{r}'_2,x_3,\ldots,x_N,t) dx_2\ldots dx_N,
\label{eq:def2RDM}
\end{multline}
where $N$ is the number of electrons. We note here, that whereas $\vec{r}_i$ represents spatial coordinates in three-dimensional space, $x_i \equiv (\vec{r}_i,\tau_i)$ denotes both the position $\vec{r}_i$ \emph{and} the spin $\tau_i$ of the $i$-th electron, and the integration symbolically also labels the summation over the spin degrees of freedom in addition to the integration over the spatial variables. In the relativistic theory with SOC, it is convenient to keep the indices associated with $\vec{r}_1$ and $\vec{r}_2$ free. As a consequence, $D$ and $\Gamma$ are still multicomponent tensors, for instance, in case of the Dirac theory, $D$ and $\Gamma$ have the dimensions of $4\times 4$ and $4\times 4\times 4\times 4$, respectively. Hence, the scalar electron charge density is obtained as
\begin{equation}
    \rho(\vec{r},t) = -e\Tr D(\vec{r};\vec{r};,t),
\label{eq:defDensity0}
\end{equation}
where $\Tr$ indicates the trace over the bispinor components. Likewise, for the four-component current density, it follows that
\begin{equation}
    \vec{j}(\vec{r},t) = -ec \Tr \vec{\alpha}D(\vec{r};\vec{r};,t).
\label{eq:defDensityCurrent}
\end{equation}
Eqs.~\eqref{eq:defDensity0} and~\eqref{eq:defDensityCurrent} generalize the one-electron definitions of the charge and current densities in Eqs.~\eqref{eq:defDensity01e} and~\eqref{eq:defDensityCurrent1e} for many-electron wave functions, since they are agnostic to the method that was used to calculate the one-electron RDM.

Exact time propagation determined by Eq.~\eqref{eq:eomexact} can equivalently be formulated in the language of RDMs, which avoids the use of the cumbersome many-electron wave function. Let us assume that we have a set of orthonormal spin-orbitals $\varphi_p(\vec{r})$. The one- and two- RDMs matrices in the spin-orbital basis then become
\begin{equation}
    D_{pq}(t) = \int d^3\vec{r}_1 \int d^3\vec{r}'_1 \dg{\varphi}_p(\vec{r}_1) D(\vec{r}_1;\vec{r}'_1;t) \varphi_q(\vec{r}'_1),
\label{eq:def1RDMmo}
\end{equation}
and
\begin{equation}
    \Gamma_{pqrs}(t) = \int d^3\vec{r}_1 \ldots d^3\vec{r}'_2 \dg{\varphi}_p(\vec{r}_1) \dg{\varphi}_r(\vec{r}_2) \Gamma(\vec{r}_1,\vec{r}_2;\vec{r}'_1,\vec{r}'_2;t) \varphi_q(\vec{r}'_1) \varphi_s(\vec{r}'_2),
\label{eq:def2RDMmo}
\end{equation}
respectively. The time-dependent one-electron RDM can be obtained by solving the equation of motion of Liouville-von Neumann (LvN) type~\cite{kadek2018advancing}
\begin{equation}
    i\hbar\pd{}{t}\mathbf{D}(t) = \left[\mathbf{h}(t),\mathbf{D}(t)\right] + \frac{1}{2} \Tr_1\left[\mathbf{G},\mathbf{\Gamma}(t)\right],
\label{eq:eom1RDM}
\end{equation}
where $[,]$ denotes the commutator, $\mathbf{h}(t)$ and $\mathbf{G}$ are matrices of one- and antisymmetrized two-electron integrals
\begin{gather}
        h_{pq}(t) \equiv \int \dg{\varphi}_p(\vec{r}) \hat{h}^{\text{D}}(t) \varphi_q(\vec{r}) d^3\vec{r}, 
        \label{eq:def1Int}
        \\
        G_{pqrs} \equiv \mathcal{I}_{pqrs} - \mathcal{I}_{psrq};
        \quad
        \mathcal{I}_{pqrs} \equiv \iint \dg{\varphi}_p(\vec{r}_1)\varphi_q(\vec{r}_1) r^{-1}_{12} \dg{\varphi}_r(\vec{r}_2)\varphi_s(\vec{r}_2) d^3\vec{r}_1 d^3\vec{r}_2, \label{eq:def2Int}
\end{gather}
and
\begin{align}
        (\mathbf{G\Gamma})_{pqrs} &\equiv G_{perf}\Gamma_{eqfs}, \label{eq:eom1RDMnotation2}\\
        (\Tr_1 \mathbf{X})_{pq} &\equiv X_{pqrr} \label{eq:eom1RDMnotation3}
\end{align}
for any two-electron matrix $\mathbf{X}$. Upon inspecting Eq.~\eqref{eq:eom1RDM}, we can see that the exact time evolution of the one-electron RDM also depends on the two-electron RDM $\mathbf{\Gamma}(t)$, which is also not known. Likewise, we could proceed by writing the equation of motion for the two-electron RDM. However, in general, the equation of motion for the $N$-electron RDM will contain the RDM of the order $N+1$, leading to an infinite hierarchy of coupled equations for RDMs, mirroring the same situation that occurs in the theory of Green's functions~\cite{fetter2003quantum}. Solving the resulting system of equations is impractical, hence, approximations to the higher-order second term that decouple the equations are sought. In the following sections, we will describe the LvN equation for the one-electron RDM where the second term containing $\mathbf{\Gamma}(t)$ is approximated in the mean-field manner using only the one-electron RDM in the framework of time-dependent Hartree--Fock theory and density functional theory.

\subsection{Time-reversal symmetry}
\label{sec:TRS}

One of the most important properties of quantum-mechanical equations of motion (and all microscopic laws) is their symmetry with respect to the reversal of time. Let us use the shorthand notation for the many-electron wave function $\Psi(t) \equiv \Psi(x_1,\ldots,x_N,t)$. Replacing $t\rightarrow -t$ in Eq.~\eqref{eq:eomexact} gives
\begin{equation}
    -i\hbar\pd{\Psi(-t)}{t} = \hat{H}(-t)\Psi(-t).
\label{eq:eomexactTR1}
\end{equation}
This equation differs from the original one in two ways. First, the Hamiltonian is expressed in the inverted time $-t$. Second, there is an extra minus sign on the left hand side of the equation. Let us assume we have an \emph{antiunitary} operator $\mathcal{K}$ that is unitary ($\dg{\mathcal{K}}\mathcal{K} = \mathbb{I}$) and antilinear
\begin{equation}
    \mathcal{K}i = -i\mathcal{K}.
\label{eq:TRonImag}
\end{equation}
Letting this operator act from the left on the Eq.~\eqref{eq:eomexactTR1}, and denoting
\begin{align}
    \bar{\Psi}(t) &:= \mathcal{K}\Psi(-t), \label{eq:TRpsi}\\
    \bar{H}(t) &:= \mathcal{K}\hat{H}(-t)\dg{\mathcal{K}}, \label{eq:TRHam}
\end{align}
we obtain
\begin{equation}
    i\hbar\pd{\bar{\Psi}(t)}{t} = \bar{H}(t)\bar{\Psi}(t).
\label{eq:eomexactTR2}
\end{equation}
In principle, this is a new equation of motion with a new solution, however, if we can assume that the Hamiltonian satisfies $\bar{H}(t) = \hat{H}(t)$, \emph{i.e.}
\begin{equation}
    \mathcal{K}\hat{H}(t) = \hat{H}(-t)\mathcal{K},
\label{eq:TRHamSym}
\end{equation}
then Eqs.~\eqref{eq:eomexact} and~\eqref{eq:eomexactTR2} represent the \emph{same} equation, for which we obtained a pair of solutions $\Psi(t)$ and $\bar{\Psi}(t)$. The condition in Eq.~\eqref{eq:TRHamSym} is known as time-reversal symmetry (TRS).

Due to the requirement in Eq.~\eqref{eq:TRonImag}, the operator $\mathcal{K}$ must \emph{at least} contain complex conjugation. This is a sufficient condition for scalar wave functions in nonrelativistic theory, where Eq.~\eqref{eq:TRHamSym} reduces to $\hat{H}^*(t) = \hat{H}(-t)$ and additionally the condition that the time-independent Hamiltonian is real-valued. However, for spinor wave functions and multicomponent relativistic theories, $\mathcal{K}$ can have a more complicated matrix form. For instance, in case of the Dirac four-component one-electron Hamiltonian, the operator $\mathcal{K}$ takes the form~\cite{Dyall2007,saue1996,komorovsky2016new}
\begin{equation}
    \mathcal{K} = -i\begin{pmatrix}
        \sigma_y & 0_2 \\
        0_2 & \sigma_y
    \end{pmatrix} \mathcal{K}_0,
\label{eq:TROperator4c}
\end{equation}
where $\mathcal{K}_0$ denotes the complex conjugation, and $\sigma_y$ is the $y$-th $2\times 2$ Pauli matrix.

We conclude this section by noting that the condition in Eq.~\eqref{eq:TRHamSym} is satisfied for nonrelativistic as well as relativistic Hamiltonians. Neither internal electromagnetic interactions nor spin--orbit coupling terms break TRS, \emph{i.e.} the Hamiltonain consists of symmetric operators ($\mathcal{K}\hat{A}\dg{\mathcal{K}}=\hat{A}$) or bilinear products of antisymmetric ($\mathcal{K}\hat{A}\dg{\mathcal{K}}=-\hat{A}$) operators, such as $\vec{\sigma}\cdot\vec{p}$, that are again symmetric. However, if external fields are introduced, they can break the TRS, for instance, an electric field with the time dependence given by an odd function of $t$. More importantly, it is often discussed in the literature that the presence of a magnetic field breaks TRS. This is only true if the magnetic field $\vec{B}$ is considered as external and does not change its orientation upon time reversal, \emph{i.e.} $\mathcal{K}$ only acts on the electronic degrees of freedom. In such situations, terms like $\vec{B}\cdot\hat{\vec{S}}$, where $S$ denotes the electron spin, become antisymmetric with respect to the time reversal, because the operator $\mathcal{K}$ only acts on $\hat{\vec{S}}$ ($\mathcal{K}\hat{\vec{S}}\dg{\mathcal{K}}=-\hat{\vec{S}}$) and not on $\vec{B}$.

\section{Equations-of-motion for approximate-state wave functions}
\label{sec:EOM4approxWF}
The previous section dealt with exact state theory. In practical calculations,
\textit{model quantum chemistries} are used to treat systems containing many particles.
The theory presented here focuses on both time-dependent Hartree–Fock (TDHF) theory
and time-dependent density functional theory (TDDFT) in the time-dependent Kohn--Sham (TDKS) framework.
From the practical point of view, both TDHF and KS TDDFT are mean-field
theories solving equations for one-electron molecular orbitals. Therefore, we use the term
time-dependent self-consistent field (TDSCF) when referring to both methods together.
In the following text, we sketch the derivation of working equations for TDHF and TDKS theories.
Since the final form of the equations is the same for both methods, the rest of this chapter
concerning propagators, evaluation of molecular properties, and analysis applies equally
to both of them.

\subsection{Time-dependent Hartree--Fock theory}

The main idea of the TDHF method is to approximate the many-electron time-dependent wave function
$\Psi(x_1,x_2,\ldots;t)$ by a single Slater determinant built from time-dependent molecular
spin-orbitals (MO) $\varphi_i (x,t)$, where we grouped the electron's spatial and spin degrees of freedom
into a single variable $x\equiv(\vec{r},\tau)$. Hence, the ansatz reads
\begin{flalign}
\label{eq:TDslaterDet}
\Psi(x_1,x_2,\ldots;t)
& =
\frac{1}{\sqrt{N!}}
\begin{vmatrix}
\varphi_1 (x_1,t) & \varphi_1 (x_2,t) & \cdots & \varphi_1 (x_N,t) \\
\varphi_2 (x_1,t) & \varphi_2 (x_2,t) & \cdots & \varphi_2 (x_N,t) \\
\vdots              & \vdots              & \ddots & \vdots              \\
\varphi_N (x_1,t) & \varphi_N (x_2,t) & \cdots & \varphi_N (x_N,t) \\
\end{vmatrix}
\nonumber \\
& =
\frac{1}{\sqrt{N!}} \sum_{\{P\}} (-1)^{|P|} \left|\varphi_{P(1)} (x_1,t) \varphi_{P(2)} (x_2,t) \ldots \varphi_{P(N)} (x_N,t) \right|
,
\end{flalign}
where $P$ denotes a permutation of indices, $P(i)$ is the new index after permutation, and $\sum_{\{P\}}$ is the sum over all possible permutations of MO indices. The prefactor $(-1)^{|P|}$ is the sign $\pm 1$ of the permutation based on the permutation length $|P|$. This ansatz uses complex spin-orbitals instead of real scalar orbitals and facilitates a direct extension of the nonrelativistic HF theory into the relativistic domain. Furthermore, in the TDHF theory, we assume that the many-electron wave function retains the form of a single
Slater-determinant during the entire time evolution.

The working equations of TDHF can be derived using the time-dependent variational principle.
Several functionals to be minimized have been formulated, such as the Dirac--Frenkel functional~\cite{frenkel1934wave,helgaker2012recent}
\begin{equation}
I^\mathrm{DF} = \int dt \Braket{\Psi(t)|i\hbar\pd{}{t} - \hat{H}|\Psi(t)},
\end{equation}
or the McLachlan functional
\begin{equation}
\label{eq:McLachlan}
I^\mathrm{ML}(t) = \Braket{\left(i\hbar\pd{}{t} - \hat{H}\right)\Psi(t)|\left(i\hbar\pd{}{t} - \hat{H}\right)\Psi(t)},
\end{equation}
where we used the braket notation $\Braket{\ldots}$ to indicate the integration over degrees of freedom (spin and spatial) of all electrons. These functionals can be used to derive the final form of the TDHF equations for MOs~\cite{Tannor2007}
\begin{equation}
\label{eq:tdhf-mo}
i\hbar \frac{\partial}{\partial t}\varphi_i(\vec{r},t)
=
\hat{F}_\text{HF}\left[\{\varphi_j(\vec{r},t)\}\right]\!(\vec{r},t)~\varphi_i(\vec{r},t)
,
\end{equation}
where $\hat{F}_\text{HF}$ is the Fock operator known from time-independent HF theory. Here, $\hat{F}_\text{HF}$ contains the Coulomb interaction of the electron with the mean-field of other electrons, the Fock exchange operator, and the one-electron Dirac Hamiltonian $\hat{h}^\text{D}(t)$ that includes the interaction with external time-dependent electromagnetic fields. As a consequence, in the four-component Dirac theory, $\hat{F}_\text{HF}$ is a $4\times 4$ operator acting on bispinor orbitals $\phi_i$. The presence of the explicitly time-dependent external fields in the one-electron part of $\hat{F}_\text{HF}$ and the dependence of the mean-field and exchange terms on the spin-orbitals, which are now time-dependent, represents the most distinct difference of the Fock operator in the TDHF theory from its time-independent counterpart.

\subsection{Time-dependent Kohn--Sham DFT}

Analogously to the static case, the idea of time-dependent density functional
theory (TDDFT)\cite{ullrich2012timedependent}
is to replace the many-electron wave function of $3N$ spatial variables and time with the simpler object --- electron density $\rho(\vec{r},t)$.
In the nonrelativistic framework, the theoretical foundations of TDDFT are provided by two theorems.  The first one,
the Runge--Gross theorem~\cite{runge1984density},
is a time-dependent analogue of the Hohenberg--Kohn
theorem connecting the time-dependent external potential and time-dependent
density. The second one, the Van Leeuwen theorem~\cite{vanLeeuwen1999mapping},
connects the real system with a fictitious system with different interaction potential.
Application of these theorems allows for introducing a fictitious KS system of non-interacting electrons for which the many-electron wave function is a single Slater determinant built of one-particle functions called KS orbitals. The final TDKS equations are similar to Eq.~\eqref{eq:tdhf-mo},
\begin{equation}
\label{eq:tdks-mo}
i\hbar \frac{\partial}{\partial t}\varphi_i(\vec{r},t)
=
\hat{F}_\text{KS}\left[\{\varphi_j(\vec{r},t)\}\right]\!(\vec{r},t)~\varphi_i(\vec{r},t)
,
\end{equation}
except that the Fock operator $\hat{F}_\text{KS}$ also contains the exchange--correlation (XC) potential derived from the approximation to the XC energy functional instead of the exact HF exchange term. This XC term links the fictitious KS system to the studied real system. In hybrid DFT~\cite{becke1993new}, functionals allow for a fraction of the HF exchange contribution to also enter $\hat{F}_\text{KS}$, bringing the very important element of (exact) antisymmetry of the many-electron wave function to DFT.

Approximating the XC functional in the time domain is more challenging than in time-independent theory.
In principle, TDDFT requires the development and use of special time-dependent XC potentials that may
generally depend on the density in previous times.
However, a widespread practice is to simply use potentials from
time-independent DFT, with the time variable only entering via the time-dependence of the density (and its gradient).
This local-in-time approximation is known as the adiabatic approximation in TDDFT~\cite{bauernschmitt1996treatment,provorse2016electron}.
The term adiabatic approximation is actually used to label a combination of two approximations: firstly the
adiabatic approximation itself~\cite{ullrich2012timedependent} and secondly the approximations that were used in the construction of the time independent XC functional~\cite{ziesche1998density}.
The adiabatic approximation is valid when a system remains in its instantaneous eigenstate for slowly varying perturbations that act on it~\cite{ullrich2012timedependent} and is widely used in TDDFT due to the lack of accepted time (memory) dependent functionals. The memory effects were also shown to be negligible in the context of nonlinear processes and strong-field excitations studied in non-perturbative electron dynamics~\cite{thiele2008adiabatic}. However, non-adiabatic effects in the XC functional become important for high-frequency oscillations~\cite{thiele2008adiabatic}, double and charge-transfer excitations~\cite{ullrich2006time}. Extending the XC potential beyond the adiabatic approximation while still exploiting the local gradient expansion can be achieved if the current density is used as a central variable~\cite{vignale1996current,ullrich2006nonadiabatic}. More general framework for time-dependent functionals with memory in TDDFT introduces viscoelastic stresses known in hydrodynamics for the electron liquid~\cite{vignale1997time,ullrich2006time} or formulates TDDFT in a comoving Lagrangian reference frame~\cite{tokatly2005quantum}.

In a similar manner, the extension of DFT to the relativistic domain~\cite{rajagopal1973inhomogeneous,saue2002four} also makes use of non-relativistic XC potentials that take relativistic densities as input. Linear-response TDDFT has been extended to the relativistic approximate two-component framework and applied to calculate absorption spectra of solids~\cite{romaniello2007relativistic}, however, proper theoretical foundations that incorporate both effects of time-dependent fields as well as relativity and generalize the Runge--Gross and Van Leeuwen theorems to the relativistic domain do not exist. Despite this, the relativistic real-time and linear-response TDDFT has been applied to study a number of molecular properties, as will be discussed in more detail in Section~\ref{sec:applications}.

\subsection{Liouville--von Neumann equation in four-component framework}

As discussed in Section~\ref{sec:RDMs}, it is often more practical to work in the formalism of density matrices. This is especially the case for theories that express the many-electron wave function as a single Slater determinant, such as TDHF and TDKS, where the two-electron RDM is not needed, and the entire information about the time evolution of a quantum state of the many-electron system is encoded in the one-electron RDM.

Let us express the time-dependent spin-orbitals $\varphi_i(\vec{r},t)$ appearing in Eqs.~\eqref{eq:tdhf-mo} and~\eqref{eq:tdks-mo} using a set of $n$ static \emph{orthonormal} functions $\{X(\vec{r})\}$. Then
\begin{equation}
  \label{eq:basisExpansion}
  \varphi_i(\vec{r},t) = \sum_{\mu}^{n}X_{\mu}(\vec{r}) C_{\mu i}(t),
\end{equation}
where $C_{\mu i}(t)$ are the complex-valued expansion coefficients. For purposes of this Chapter, $\{X(\vec{r})\}$ shall refer to orthonormal atomic orbitals (AOs). For cases where the wave function $\Psi(t)$ is a Slater determinant, the one-electron RDM from Eq.~\eqref{eq:def1RDM} can be expressed through the occupied (occ) spin orbitals as
\begin{equation}
\label{eq:1RDMinTDSCF}
    D(\vec{r};\vec{r}';t) = \sum_i^\text{occ} \varphi_i(\vec{r},t)\dg{\varphi}_i(\vec{r}',t).
\end{equation}
Inserting Eqs.~\eqref{eq:basisExpansion} and~\eqref{eq:1RDMinTDSCF} into Eq.~\eqref{eq:def1RDMmo} gives the following orthonormal AO representation of the RDM
\begin{equation}
\label{eq:1RDMinTDSCF}
    D_{\mu\nu}(t) = \sum_i^\text{occ} C_{\mu i}(t)\dg{C}_{\nu i}(t).
\end{equation}
Introducing the matrices $\mathbf{D}(t)$ and $\mathbf{C}(t)$ with elements $D_{\mu\nu}(t)$ and $C_{\mu i}$, respectively, we can write
\begin{equation}
\label{eq:1RDMinTDSCF}
    \mathbf{D}(t) = \mathbf{C}(t)\dg{\mathbf{C}}(t).
\end{equation}
Taking the time derivative of this equation and using the time-dependent equations for spin-orbitals (Eqs.~\eqref{eq:tdhf-mo} or~\eqref{eq:tdks-mo}) in combination with the expansion in Eq.~\eqref{eq:basisExpansion}, we obtain the Liouville--von Neumann (LvN) equation of motion (EOM) for the RDM
\begin{equation}
\label{eq:LvN0}
i\hbar\frac{\partial \mathbf{D}(t)}{\partial t} 
= 
[\mathbf{F}(t),\mathbf{D}(t)]
,
\end{equation}
where we dropped the HF and KS labels on the Fock matrix $\mathbf{F}(t)$. We note, that in the HF theory, this equation coincides with the general Eq.~\eqref{eq:eom1RDM}, since the two-electron RDM is approximated as
\begin{equation}
\label{eq:twoRDMfromOneRDM}
    \Gamma_{\mu\nu\kappa\lambda}(t) = D_{\mu\nu}(t)D_{\kappa\lambda}(t) - D_{\mu\lambda}(t)D_{\kappa\nu}(t).
\end{equation}
Using this factorization of the two-electron RDM in Eq.~\eqref{eq:eom1RDM} gives rise to both the mean-field Coulomb as well as the exact exchange terms that complement the one-electron Hamiltonian in the Fock matrix of HF theory. The advantage of solving the LvN equation over the respective equations for spin orbitals is that the RDM is gauge invariant with respect to orbital rotations $\varphi'_p \rightarrow \varphi'_q V_{qp}$, \emph{i.e.} unitary transformations $\mat{V}$ that do not mix the occupied and virtual spin orbitals. This gauge freedom of the orbitals was utilized in the work of Jia \emph{et al.}~\cite{jia2018fast} to allow for much larger time steps used in real-time simulations based on solving the EOM for orbitals in the parallel transport gauge.

Real-time methods are based on directly solving Eq.~\eqref{eq:LvN0} in the time domain by numerically propagating the RDM (see Section~\ref{sec:propagation}). Since the Fock matrix is Hermitian, the time evolution must be unitary. However, Eq.~\eqref{eq:LvN0} is sometimes augmented by an extra term to model the relaxation of the system to the equilibrium (ground) state $\mat{D}_{eq}$ with an empirical rate of relaxation matrix $\gamma$. The LvN equation then reads
\begin{equation}
\label{eq:LvN}
i\hbar\frac{\partial \mathbf{D}(t)}{\partial t} 
= 
[\mathbf{F}(t),\mathbf{D}(t)]
- 
i\hbar\mat{\gamma} \big( \mat{D}(t) - \mat{D}_{eq} \big)
.
\end{equation}
At the level of theory presented here, the matrix $\gamma$ is phenomenological and is commonly approximated by a single parameter, referred to as a damping parameter. In this case, the LvN equation can be solved without the damping parameter and its application is postponed to a post-processing step (see discussion below). We note, that if the damping term is included in the LvN equation, the time propagation is no longer unitary and the energy of the system is not conserved even in the absence of external field(s). 

Within a finite time window, the solution of the LvN equation Eq.~\eqref{eq:LvN0} reduces to the evaluation of the time-dependent Fock matrix at discrete time steps, and to the propagation of the density matrix in time. Here, we briefly outline the main features of the Fock matrix evaluation, assuming the full four-component level of theory. By following the previous discussion, the Fock matrix in Eq.~\eqref{eq:LvN0} is given in a set of $n$ \emph{orthonormal} AOs $\{X(\vec{r})\}$,
\begin{equation}
    F_{\mu\nu}(t)
    =
    \Braket{X_{\mu}(\vec{r})|\hat{F}\left[\{\varphi(\vec{r},t)\}\right]\!(\vec{r},t)|X_{\nu}(\vec{r})}
    .
\end{equation}
Of particular interest in this Chapter are applications where molecular systems are irradiated by classical time-dependent electric field(s). In this case,
the 4c Fock matrix can easily be derived from the electronic Hamiltonian in the 
length-dipole representation (see Eq.~\eqref{eq:hint-length+dipole} in Section~\ref{sec:relatHamiltonian})~\cite{repisky2015excitation,moitra2023accurate}
 \begin{align}
  \label{eq:4cFock}
  F^{\text{4c}}_{\mu\nu}(t)
  =
  F^{\mathrm{4c}}_{\mu\nu}[\boldsymbol{\mathcal{E}},\boldsymbol{\mathcal{F}}](t)
  & =
  h^{\mathrm{D}}_{\mu\nu}
  +
  \sum_{\kappa\lambda}^{n}
  G^{\text{4c}}_{\mu\nu,\kappa\lambda}
  D^{\text{4c}}_{\lambda\kappa}(t,\boldsymbol{\mathcal{E}},\boldsymbol{\mathcal{F}})
  \\[0.2cm] & 
  +
  \sum_{u\in0,x,y,z}
  \int v^{xc}_{u}\!\left[\boldsymbol{\rho}^\text{4c}(\vec{r},t,\boldsymbol{\mathcal{E}},\boldsymbol{\mathcal{F}})\right]
  \Omega_{u,\mu\nu}^{\text{4c}}(\vec{r})\,d^{3}\vec{r}
  \nonumber
  \\[0.2cm] &
  -
  \sum_{u\in x,y,z}
  P^{\mathrm{4c}}_{u,\mu\nu}
  \mathcal{E}_u(t)
  -
  \sum_{u\in x,y,z}
  P^{\mathrm{4c}}_{u,\mu\nu}
  \mathcal{F}_u(t)
  \nonumber
  .
\end{align}
The right-hand side includes the matrix representation of the one-electron Dirac operator, the two-electron (2e) Coulomb interaction operator, the exchange--correlation (xc) operator, and the particle--field interaction operators. For generality we involve two time-dependent electric fields $\bm{\mathcal{E}}(t)$ and $\bm{\mathcal{F}}(t)$ which are coupled to the molecular system via the electric dipole moment operator matrix ($\mathbf{P}^{\mathrm{4c}}_{u}$). 

Computationally most demanding is the 2e contribution as it requires the evaluation of generalized anti-symmetrized electron repulsion integrals (ERIs)~\cite{ReSpect}
\begin{equation} 
   \label{eq:eri}
   G^{\text{4c}}_{\mu\nu,\kappa\lambda}
    =
   \mathcal{I}^{\text{4c}}_{\mu\nu,\kappa\lambda}
   -
   \zeta\mathcal{I}^{\text{4c}}_{\mu\lambda,\kappa\nu}
   ;\qquad
   \mathcal{I}^{\text{4c}}_{\mu\nu,\kappa\lambda}
   =
   \iint
   \Omega_{0,\mu\nu}^{\text{4c}}(\vec{r}_{1})
   r_{12}^{-1}
   \Omega_{0,\kappa\lambda}^{\text{4c}}(\vec{r}_{2})
  d^{3}\vec{r}_{1}d^{3}\vec{r}_{2}
  ,
\end{equation}
in terms of 4c \emph{charge} distribution functions
\begin{equation} 
   \label{eq:omega0}
   \Omega^{\text{4c}}_{0,\mu\nu}(\vec{r})
   =
   X_{\mu}^{\dagger}(\vec{r}) 
   X_{\nu}(\vec{r})
  . 
\end{equation}
Here, each 4c basis function $X_{\mu}(\vec{r})\equiv\{X^{\text{L}}_{\mu}(\vec{r})\oplus X^{\text{S}}_{\mu}(\vec{r})\}$ consists of the direct product of the large 2c function $X^{\text{L}}_{\mu}(\vec{r})$ and the small 2c function $X^{\text{S}}_{\mu}(\vec{r})$, related to each other to the lowest order in $c^{-1}$ by the restricted kinetically balanced (RKB) relation~\cite{Stanton1984}: $X^{\text{S}}_{\mu}\simeq(\vec{\sigma}\cdot\vec{p}X^{\text{L}}_{\mu})$. Obvious computational cost and complexity of 4c ERIs arise from the presence of the canonical momentum operator ($\vec{p}$) as well as the Pauli spin operator ($\vec{\sigma}$) in the expression for the small-component basis. Therefore, as discussed in Ref.~\cite{ReSpect}, a single 4c ERI requires even in the most compact formalism of real quaternions the simultaneous evaluation and processing of 25 times more real scalar integrals than the simpler 1c or 2c cases. This ratio further increases when RKB is substituted by the restricted magnetically balanced (RMB) relation~\cite{Komorovsky2008,Repisky2009}, which is needed for handling interactions with magnetic fields and requires the ERI evaluation formalism to be based on complex quaternions~\cite{ReSpect}.   

In addition to the charge distribution function $\boldsymbol{\Omega}_0^{\text{4c}}(\vec{r})$ used in Eq.~\eqref{eq:eri}, one can define three \emph{spin} distribution functions along the Cartesian directions
\begin{equation} 
   \label{eq:omega1-3}
   \Omega^{\text{4c}}_{k,\mu\nu}(\vec{r})
   =
   X_{\mu}^{\dagger}(\vec{r}) 
   \Sigma_k 
   X_{\nu}(\vec{r})
   ;
   \qquad
   \Sigma_{k} =
   \begin{pmatrix}
       \sigma_{k} & 0_2 \\
       0_2  & \sigma_{k}
   \end{pmatrix}
   ;
   \qquad
   k\in x,y,z
   ,
\end{equation}
in terms of which the 4c electron charge density ($\rho_0^{\text{4c}}$) as well as the electron spin densities ($\rho_{x}^{\text{4c}},\rho_{y}^{\text{4c}},\rho_{z}^{\text{4c}}$) have a particularly simple form
\begin{equation} 
   \label{eq:ng:4c}
   \rho_k^{\text{4c}} 
   =
   \rho_k^{\text{4c}}(\vec{r},t) 
   =
   \sum_{\mu\nu}^{n}
   \Omega_{k,\mu\nu}^{\text{4c}}(\vec{r}) D^{\text{4c}}_{\nu\mu}(t)
   ;\qquad
   k\in0,x,y,z
   ,
\end{equation}
where $\Sigma_k$ is the Dirac spin operator. Note that all current \emph{noncollinear} extensions of nonrelativistic xc functionals employ those four densities (alongside of their gradients) as basic variables~\cite{Kubler1988,Sandratskii1998,VanWuellen2002,Scalmani2012,Komorovsky2019}. In the relativistic 2c and 4c theory, the use of a noncollinear formalism is necessary since the spatial and spin degrees of freedom are no longer independent and are coupled by the spin-orbit interaction. This coupling results in a lack of rotational invariance of the xc energy if the energy is calculated collinearly through the $z$ spin-component only~\cite{VanWuellen2002}. A common way to circumvent this variance problem is to formulate the nonrelativistic exchange–correlation functionals noncollinearly. Therefore, we utilize in our real-time TDSCF implementation the noncollinear variables of Scalmani and Frisch~\cite{Scalmani2012} and evaluate the noncollinear xc potential $v_k^{xc}$ in Eq.~\eqref{eq:4cFock} within a generalized gradient approximation as
\begin{equation}
   \label{eq:vxc:4c}
   v^{xc}_k\!\left[\boldsymbol{\rho}^{\text{4c}}(t)\right]
   =
   \frac{\partial\varepsilon^{xc}}{\partial \rho_k^{\text{4c}}(t)}
   -
   \left( \boldsymbol{\nabla} \cdot
          \frac{\partial\varepsilon^{xc}}{\partial \boldsymbol{\nabla}\rho_k^{\text{4c}}(t)}\right)
   ;\qquad
   k\in0,x,y,z
   .
\end{equation}
Here, $\varepsilon^{xc}$ and $\vec{\rho}^{\text{4c}}$ refer to a nonrelativistic xc energy density and an electron density vector consisting of the electron charge and spin densities (together with their gradients). For further details on our noncollinearity implementation, the reader is referred to Refs.~\cite{ReSpect,Komorovsky2019}.


\subsection{Reduction of the Liouville--von Neumann equation to the exact two-component (X2C) form}

While full four-component (4c) relativistic real-time electron dynamics simulations are nowadays feasible~\cite{repisky2015excitation,Kadek2015,Konecny2016,moitra2023accurate,Konecny-JCP-149-204104-2018,DeSantis2020pyberthart}, there is interest in developing approximate methods enabling these simulations to be performed more efficiently at the two-component (2c) level while maintaining the accuracy of the parent 4c regime. Therefore, we shall discuss the transformation of the original 4c Liouville-von Neumann (LvN) equation to its 2c form, with a particular focus on the modern exact two-component (X2C) formalism. 

The X2C Hamiltonian model has gained wide popularity in recent years as it reduces the original 4c problem by half while requiring only a few simple algebraic manipulations~\cite{Heully1986,Jensen2005,Kutzelnigg2005,Liu2007,Ilias2007}. However, 
accuracy of this Hamiltonian strongly depends on the two-electron (2e) and exchange–correlation (xc) picture-change correction models employed~\cite{Knecht2022} and can vary as much as 5-6 orders of magnitude for core-shell energies. Since the pioneering X2C RT-TDDFT implementations~\cite{Konecny2016,Goings2016} utilize a crude one-electron X2C (1eX2C) Hamiltonian model where the picture-change corrections are entirely neglected, the inner-shell spinors (and their energies) substantially differ from the reference 4c results~\cite{Knecht2022}. Therefore, our focus here is to provide theoretical insights into three numerically accurate X2C Hamiltonian models~\cite{Knecht2022,moitra2023accurate,Konecny2023}, dubbed as amfX2C, eamfX2C and mmfX2C that enable accounting for the two-electron and exchange-correlation picture-change effects.

By following the matrix-algebraic approach of X2C, let us assume that at an arbitrary time $t$ there exists a unitary transformation matrix $\mathbf{U}(t)$ that block-diagonalizes/decouples the 4c Fock matrix
\begin{equation}
    \label{x4c:Fock}
    \mathbf{F}^{\text{4c}}(t)
    \rightarrow
    \tilde{\mathbf{F}}^{\text{4c}}(t)
    =
    \mathbf{U}^{\dagger}(t)\mathbf{F}^{\text{4c}}(t)\mathbf{U}(t)
    =
    \left(
    \begin{array}[c]{cc}
      \mathbf{\tilde{F}}^\mathrm{LL}(t) & \bm{0}_{2}      \\
      \bm{0}_{2} & \mathbf{\tilde{F}}^\mathrm{SS}(t)      \\
    \end{array}
    \right)
    .
\end{equation}
Note that: (i) we use tildes to indicate all transformed quantities; (ii) $\mathbf{F}^{\text{4c}}(t)$ and $\mathbf{U}(t)$ also depend on the electric field
$\boldsymbol{\mathcal{E}}(t)$ and $\boldsymbol{\mathcal{F}}(t)$, but for clarity of presentation this dependence is omitted now.
Under the X2C transformation, the parent 4c EOM for MO coefficients becomes
\begin{equation}
  \label{x4c:Ceom}
  i\hbar
  \frac{\partial\boldsymbol{\tilde{C}}^\mathrm{4c}_{i}(t)}{\partial t}
  = 
  \mathbf{\tilde{F}}^\mathrm{4c}(t)\boldsymbol{\tilde{C}}^\mathrm{4c}_{i}(t)
  +
  i\hbar\left(\frac{\partial\mathbf{U}^\dagger(t)}{\partial t}\right) \mathbf{U}(t) \boldsymbol{\tilde{C}}^\mathrm{4c}_{i}(t)
  ,
\end{equation}
where
\begin{equation}
    \label{x4c:C}
    \boldsymbol{\tilde{C}}^\mathrm{4c}_{i}(t)
    =
    \mathbf{U}^{\dagger}(t)\boldsymbol{C}^\mathrm{4c}_{i}(t)
    .
\end{equation}
A similar relation also holds for the X2C transformed LvN equation
\begin{equation}
  \label{x4c:LvN0}
  i\hbar\frac{\partial\mathbf{\tilde{D}}^\mathrm{4c}(t)}{\partial t}
  = 
  \left[\mathbf{\tilde{F}}^\mathrm{4c}(t), \mathbf{\tilde{D}}^\mathrm{4c}(t)\right]
  +
  i\hbar\left[ \left(\frac{\partial\mathbf{U}^\dagger(t)}{\partial t}\right) \mathbf{U}(t), \mathbf{\tilde{D}}^\mathrm{4c}(t) \right]
  ,
\end{equation}
with the density matrix
\begin{equation}
      \tilde{\mathbf{D}}^{\text{4c}}(t)
      =
      \sum_{i}^{\text{occ}}
      \boldsymbol{\tilde{C}}^\mathrm{4c}_{i}(t)\left(\boldsymbol{\tilde{C}}^\mathrm{4c}_{i}(t)\right)^{\dagger}
      =
      \mathbf{U}^{\dagger}(t)\mathbf{D}^{\text{4c}}(t)\mathbf{U}(t)
      . 
\end{equation}
The right hand side of Eqs.~\eqref{x4c:Ceom} and \eqref{x4c:LvN0} involves
the matrix product $\mathbf{\dot{U}}^\dagger(t)\mathbf{U}(t)$ which has nonzero off-diagonal blocks that prevent expressing these equations in the complete decoupled (block-diagonal) form. However, as discussed in Ref.~\cite{Konecny2023} for the case of a single electric field $\boldsymbol{\mathcal{E}}(t)$, the matrix values of $\mathbf{\dot{U}}^\dagger(t)$ are of the order $\mathcal{O}(|\boldsymbol{\mathcal{E}}|\omega c^{-1})$, and therefore become negligibly small within a weak-field limit ($|\boldsymbol{\mathcal{E}}|\!\ll\!1$) and a dipole approximation ($r \omega c^{-1}\!\ll\!1$). As a result, the X2C transformation matrix remains approximately constant in time, \emph{i.e.} $\mathbf{U}(t)\approx\mathbf{U}$, and Eqs.~\eqref{x4c:Ceom} and \eqref{x4c:LvN0} reduce to the simple form
\begin{equation}
  \begin{split}
  \label{x4c:EOMs-with-U0}
  i\hbar\frac{\partial\boldsymbol{\tilde{C}}^\mathrm{4c}_{i}(t)}{\partial t}
  = 
  \mathbf{\tilde{F}}^\mathrm{4c}(t)\boldsymbol{\tilde{C}}^\mathrm{4c}_{i}(t)
  ;\qquad
  i\hbar\frac{\partial\mathbf{\tilde{D}}^\mathrm{4c}(t)}{\partial t}
  = 
  \left[\mathbf{\tilde{F}}^\mathrm{4c}(t), \mathbf{\tilde{D}}^\mathrm{4c}(t)\right]
  .
  \end{split}
\end{equation}
This time-independence of matrix $\mathbf{U}$ is generally denoted as the \emph{adiabatic X2C transformation}~\cite{Konecny2016,Konecny2023}. 

The best possible transformation matrix $\mathbf{U}$ can be obtained from a so-called mmfX2C approach~\cite{Sikkema2009}. In this approach, $\mathbf{U}$ is obtained \emph{a posteriori} from converged 4c SCF HF/KS solutions (MO coefficients) applying for instance the one-step X2C transformation of Ilias and Saue~\cite{Ilias2007}. From the real-time dynamics point of view, these solutions are associated with the initial simulation time $t_{0}$. An important observation is that at $t_0$ the 4c occupied positive-energy MO coefficients $\vec{C}^{\text{4c}}_{i}$ as well as the 4c density matrix $\mathbf{D}^{\text{4c}}$ can be expressed in terms of their 2c counterparts,
\begin{equation}
   \begin{split}
   \label{eq:CD-transformation}
   \vec{C}^{\text{4c}}_{i}(t_0)
   &=
   \mathbf{U} \vec{\tilde{C}}^{\text{2c}}_{i}(t_0)
   \quad\Rightarrow\quad
   \big[ C^{\text{4c}} \big]^{\text{X}}_{\mu i}
   =
   \sum_{\nu}
   \big[ U \big]^{\text{XL}}_{\mu\nu} 
   \big[ \tilde{C}^{\text{2c}} \big]_{\nu i}
   \\[0.2cm]
   \mathbf{D}^{\text{4c}}(t_0)
   &=
   \mathbf{U} \mathbf{\tilde{D}}^{\text{2c}}(t_0) \mathbf{U}^{\dagger} 
   \quad\Rightarrow\quad
   \big[ D^{\text{4c}} \big]^{\text{XY}}_{\mu\nu}
   =
   \sum_{\kappa\lambda}
   \big[ U \big]^{\text{XL}}_{\mu\kappa} 
   \big[ \tilde{D}^{\text{2c}} \big]_{\kappa\lambda}
   \big[ U^{\dagger} \big]^{\text{LY}}_{\lambda\nu}
   .
   \end{split}
\end{equation}
Here, X and Y refer to the large-component (L) and small-component (S) subset of orthonormal AO basis. Within the adiabatic X2C transformation it is assumed that the relation~\eqref{eq:CD-transformation} remain valid also at an arbitrary future time $t>t_{0}$, and therefore 4c real-time dynamic results can be obtained just from the solution of simple 2c EOMs
\begin{equation}
  \label{x2c:EOMs-with-U0}
  i\hbar\frac{\partial\boldsymbol{\tilde{C}}^\mathrm{2c}_{i}(t)}{\partial t}
  = 
  \mathbf{\tilde{F}}^\mathrm{2c}(t)\boldsymbol{\tilde{C}}^\mathrm{2c}_{i}(t)
  ;\qquad
  i\hbar\frac{\partial\mathbf{\tilde{D}}^\mathrm{2c}(t)}{\partial t}
  = 
  \left[\mathbf{\tilde{F}}^\mathrm{2c}(t),\mathbf{\tilde{D}}^\mathrm{2c}(t)\right]
  .
\end{equation}
However, as shown by Knecht and coworkers for static SCF case~[\citenum{Knecht2022}], the correctly transformed 2c Fock matrix $\mathbf{\tilde{F}}^{\text{2c}}$ involves a so-called picture-change transformation of density matrix, overlap distribution matrix, and one- and two-electron integrals. Repisky and coworkers extended this observation to the time domain and derive~\cite{moitra2023accurate,Konecny2023}:
\begin{align}
  \label{eq:x2cFock}
  \tilde{F}^{\mathrm{2c}}_{\mu\nu}(t)
  =
  \left[\mathbf{U}^{\dagger}\mathbf{F}^{\text{4c}}(t)\mathbf{U}\right]^{\text{LL}}_{\mu\nu}
  & =
  \tilde{h}^{\mathrm{2c}}_{\mu\nu}
  +
  \sum_{\kappa\lambda}
  \tilde{G}^{\text{2c}}_{\mu\nu,\kappa\lambda}
  \tilde{D}^{\text{2c}}_{\lambda\kappa}(t,\boldsymbol{\mathcal{E}},\boldsymbol{\mathcal{F}})
  \\[0.2cm] & 
  +
  \sum_{u\in0,x,y,z}
  \int v^{xc}_{u}\!\left[\boldsymbol{\tilde{\rho}}^\text{2c}(\vec{r},t,\boldsymbol{\mathcal{E}},\boldsymbol{\mathcal{F}})\right]
  \tilde{\Omega}_{u,\mu\nu}^{\text{2c}}(\vec{r})\,d^{3}\vec{r}
  \nonumber
  \\[0.2cm] &
  -
  \sum_{u\in x,y,z}
  \tilde{P}^{\mathrm{2c}}_{u,\mu\nu}
  \mathcal{E}_u(t)
  -
  \sum_{u\in x,y,z}
  \tilde{P}^{\mathrm{2c}}_{u,\mu\nu}
  \mathcal{F}_u(t)
  \nonumber
  .
\end{align}
There are two important points to note here: (i) all transformed quantities are marked with tilde; (ii) the presence of the picture-change transformed charge distribution matrix ($\boldsymbol{\tilde{\Omega}}^{\text{2c}}$) in both 2e and xc interaction terms makes the evaluation of $\mathbf{\tilde{F}}^{\mathrm{2c}}$ computationally more demanding than the original 4c Fock matrix.  

Therefore, it is desirable to seek for an approximation that enables us to carry out electron dynamics simulations in 2c mode such that they are computationally efficient and reproduce the reference 4c results as closely as possible. Keeping this in mind, one can compare Eq.~\eqref{eq:x2cFock} with an approximate and computationally efficient form of the Fock matrix built with \textit{untransformed} (without the tilde) two-electron integrals $\mathbf{G}^{\text{2c}}$ and
overlap distribution matrix $\boldsymbol{\Omega}^{\text{2c}}$; that is
\begin{align}
  \label{eq:x2cFock-no2ePC}
  F^{\mathrm{2c}}_{\mu\nu}(t)
  & =
  \tilde{h}^{\mathrm{2c}}_{\mu\nu}
  +
  \sum_{\kappa\lambda}
         G^{\text{2c}}_{\mu\nu,\kappa\lambda}
  \tilde{D}^{\text{2c}}_{\lambda\kappa}(t,\boldsymbol{\mathcal{E}},\boldsymbol{\mathcal{F}})
  \\[0.2cm] & 
  +
  \sum_{u\in0,x,y,z}
  \int v^{xc}_{u}\!\left[\boldsymbol{\rho}^\text{2c}(\vec{r},t,\boldsymbol{\mathcal{E}},\boldsymbol{\mathcal{F}})\right]
  \Omega_{u,\mu\nu}^{\text{2c}}(\vec{r})\,d^{3}\vec{r}
  \nonumber
  \\[0.2cm] &
  -
  \sum_{u\in x,y,z}
  \tilde{P}^{\mathrm{2c}}_{u,\mu\nu}
  \mathcal{E}_u(t)
  -
  \sum_{u\in x,y,z}
  \tilde{P}^{\mathrm{2c}}_{u,\mu\nu}
  \mathcal{F}_u(t)
  \nonumber
  .
\end{align}
Here, it is important to emphasize that $\boldsymbol{\rho}^\text{2c}$ also remains untransformed in the sense that an untransformed $\boldsymbol{\Omega}_u^{\text{2c}}$ is used but with the correctly transformed density matrix $\mathbf{\tilde{D}}^{\text{2c}}$. We immediately find
that the difference between these two Fock matrices expresses the picture-change corrections associated with the two-electron integrals and the xc contribution
\begin{equation} 
   \label{eq:DeltaF2}
   \Delta\tilde{F}^{\text{2c}}_{\mu\nu}(t)
   =
   \tilde{F}^{\text{2c}}_{\mu\nu}(t)
   -
   F^{\text{2c}}_{\mu\nu}(t)
   =
   \sum_{\kappa\lambda}
   \Delta\tilde{G}^{\text{2c}}_{\mu\nu,\kappa\lambda}
   \tilde{D}^{\text{2c}}_{\lambda\kappa}(t)
   +
   \Delta\tilde{F}^{\text{2c,xc}}_{\mu\nu}(t)
   ,
\end{equation}
where
\begin{equation}
   \label{eq:DeltaF2-2}
   \begin{split}
   \Delta\tilde{G}^{\text{2c}}_{\mu\nu,\kappa\lambda}
   &= 
   \tilde{G}^{\text{2c}}_{\mu\nu,\kappa\lambda}
   -
   G^{\text{2c}}_{\mu\nu,\kappa\lambda},
   \\[0.2cm]
   \Delta\tilde{F}^{\text{2c,xc}}_{\mu\nu}(t)
   &= 
   \int v_k^{xc}\!\left[\boldsymbol{\tilde{\rho}}^\text{2c}(\vec{r},t)\right]
   \tilde{\Omega}_{k,\mu\nu}^{\text{2c}}(\vec{r})\,d^{3}\vec{r}
   -
   \int v_k^{xc}\!\left[\boldsymbol{\rho}^\text{2c}(\vec{r},t)\right]
   \Omega_{k,\mu\nu}^{\text{2c}}(\vec{r})\,d^{3}\vec{r}
   .
   \end{split}
\end{equation}
Here, we dropped the dependence on $\boldsymbol{\mathcal{E}}$ and $\boldsymbol{\mathcal{F}}$ for clarity. The central idea of X2C real-time electron dynamics is the solution of the 2c LvN equation~\eqref{x2c:EOMs-with-U0} with the Fock matrix
\begin{align}
  \label{eq:x2cFock-2ePC}
  \tilde{F}^{\mathrm{2c}}_{\mu\nu}(t)
  & =
  \tilde{h}^{\mathrm{2c}}_{\mu\nu}
  +
  \Delta \tilde{F}^{\text{2c}}_{\mu\nu}(t)
  +
  \sum_{\kappa\lambda}
         G^{\text{2c}}_{\mu\nu,\kappa\lambda}
  \tilde{D}^{\text{2c}}_{\lambda\kappa}(t,\boldsymbol{\mathcal{E}},\boldsymbol{\mathcal{F}})
  \\[0.2cm] & 
  +
  \sum_{u\in0,x,y,z}
  \int v^{xc}_{u}\!\left[\boldsymbol{\rho}^\text{2c}(\vec{r},t,\boldsymbol{\mathcal{E}},\boldsymbol{\mathcal{F}})\right]
  \Omega_{u,\mu\nu}^{\text{2c}}(\vec{r})\,d^{3}\vec{r}
  \nonumber
  \\[0.2cm] &
  -
  \sum_{u\in x,y,z}
  \tilde{P}^{\mathrm{2c}}_{u,\mu\nu}
  \mathcal{E}_u(t)
  -
  \sum_{u\in x,y,z}
  \tilde{P}^{\mathrm{2c}}_{u,\mu\nu}
  \mathcal{F}_u(t)
  \nonumber
  ,
\end{align}
where $\Delta\mathbf{\tilde{F}}^{\mathrm{2c}}(t)$ accounts for the picture-change corrections associated with the 2e integrals and the xc contribution.
Note that $\mathbf{\tilde{F}}^{\mathrm{2c}}(t)$ in Eqs.~\eqref{eq:x2cFock} and \eqref{eq:x2cFock-2ePC} are equal, and all differences between various flavours of X2C are due to approximations in $\Delta\mathbf{\tilde{F}}^{\mathrm{2c}}(t)$.

In the simplest but least accurate case, dubbed one-electron X2C (1eX2C), $\Delta\mathbf{\tilde{F}}^{\mathrm{2c}}(t)$ in Eq.~\eqref{eq:x2cFock-2ePC}
is completely discarded, while the decoupling matrix $\mathbf{U}$ is obtained simply from the parent one-electron Dirac Hamiltonian. This approach was employed in pioneering X2C RT-TDDFT implementations~\cite{Konecny2016,Goings2016}.    
Due to its simplicity the 1eX2C Hamiltonian still remains very popular, but caution is needed when applying this model beyond valence electric properties as shown for instance in Ref.~\cite{Konecny2023}. 

In the second model, coined as molecular mean-field X2C (mmfX2C), $\Delta\mathbf{\tilde{F}}^{\mathrm{2c}}(t)$ in Eq.~\eqref{eq:x2cFock-2ePC} is approximated
by a \emph{static} model $\Delta\mathbf{\tilde{F}}^{\mathrm{2c}}$, which is evaluated according to Eqs.~\eqref{eq:DeltaF2} and \eqref{eq:DeltaF2-2} only \emph{once} using the converged 4c molecular self-consistent field solutions~\cite{moitra2023accurate,Konecny2023}. Similarly, $\mathbf{U}$ is determined from the same 4c solutions. For theoretical and numerical justification of the static approximation used in the real-time and response mmfX2C theory, readers are referred to the original publication~\cite{Konecny2023}. Due to the late X2C transformation (post-SCF), the mmfX2C approach was found as most accurate among all X2C Hamiltonian models~\cite{Konecny2023,Knecht2022}; though, the price for this accuracy is the implementation and execution of 4c molecular SCFs. 

In line with the idea of Knecht \emph{et al.}~\cite{Knecht2022} on the amfX2C Hamiltonian for time-independent Hartree-Fock and Kohn-Sham mean-field theories, one may exploit the local atomic nature of the static picture-change correction matrix
$\Delta\mathbf{\tilde{F}}^{\text{2c}}$ discussed in the previous paragraph. In the third model, dubbed as atomic mean-field X2C (amfX2C)~\cite{moitra2023accurate,Konecny2023,Knecht2022}, 
$\Delta\mathbf{\tilde{F}}^{\mathrm{2c}}(t)$ in Eq.~\eqref{eq:x2cFock-2ePC}
is approximated by a \emph{static} model $\Delta\mathbf{\tilde{F}}^{\mathrm{amfX2C}}_{\bigoplus}$ obtained by a superposition of converged atomic quantities rather than the converged molecular one, \emph{i.e.}
\begin{align}
  \Delta\mathbf{\tilde{F}}^{\text{2c}}(t)
  \approx
  \Delta\mathbf{\tilde{F}}^{\text{amfX2C}}_{\bigoplus}
  =
  \bigoplus_{K=1}^{M}
  \Delta\mathbf{\tilde{F}}^{\text{2c}}_{K}[\mathbf{\tilde{D}}_K^{\text{2c}}]
  .
\end{align}
Here, $K$ runs over all atoms in an $M$-atomic system.

The main advantage of the amfX2C approach is that it introduces picture-change corrections to both spin-independent and spin-dependent parts of the two-electron and xc interaction just from simple atomic quantities. On the other hand, the fact that
$\Delta\mathbf{\tilde{F}}^{\text{2c}}_{\bigoplus}$ has only atomic diagonal blocks means that, for instance, the off-diagonal electron-nucleus contribution
will not cancel out with the direct electron-electron contribution at long distances from the atomic centers. This becomes problematic in solid-state calculations, where the exact cancellation of these contributions is essential at long distances. In fact, this motivated Knecht and coworkers~\cite{Knecht2022} to introduce our last X2C Hamiltonian model, called extended amfX2C (eamfX2C). The generalization of eamfX2C to the time domain was recently discussed by Konecny and coworkers~\cite{Konecny2023}, and it requires to approximate
$\Delta\mathbf{\tilde{F}}^{\mathrm{2c}}(t)$ in Eq.~\eqref{eq:x2cFock-2ePC} by a \emph{static} model $\Delta\mathbf{\tilde{F}}^{\mathrm{eamfX2C}}_{\bigoplus}$ obtained from the time-independent version of equation~\eqref{eq:DeltaF2}
\begin{equation} 
   \Delta\tilde{F}^{\text{eamfX2C}}_{\bigoplus,\mu\nu}
   =
   \tilde{F}^{\text{2c}}_{\bigoplus,\mu\nu}
   -
   F^{\text{2c}}_{\bigoplus,\mu\nu}
   =
   \sum_{\kappa\lambda}
   \Delta\tilde{G}^{\text{2c}}_{\mu\nu,\kappa\lambda}
   \tilde{D}^{\text{2c}}_{\bigoplus,\lambda\kappa}
   +
   \Delta\tilde{F}^{\text{2c,xc}}_{\bigoplus,\mu\nu}
   ,
\end{equation}
with elements on the right hand side given in Eq.~\eqref{eq:DeltaF2-2}. The picture-change corrections associated with the two-electron integrals and the xc contribution involve the 2c density matrix $\mathbf{\tilde{D}}^{\text{2c}}_{\bigoplus}$
obtained from a superposition of converged 4c atomic density matrices $\mathbf{D}^{\text{4c}}_{K}$, i.e.
\begin{align}
  \mathbf{\tilde{D}}^{\text{2c}}_{\bigoplus}
  =
  \bigoplus_{K=1}^{M}
  \left[
  \mathbf{U}_{K}^{\dagger}
  \mathbf{D}^{\text{4c}}_{K}
  \mathbf{U}_{K}
  \right]^{\text{LL}}
  .
\end{align}
Here, $K$ runs over all atoms in an $M$-atomic system.

\section{Real-time propagation}
\label{sec:propagation}

\subsection{Evolution operator}

The solution of the TDHF and TDKS equations for an arbitrary time $t$ can be written in the compact form by defining the \emph{evolution operator} $U(t,t')$ that propagates the state from time $t'$ to time $t$ as
\begin{equation}
     \varphi_i(\vec{r},t) = U(t, t') \varphi_i(\vec{r},t')
     ,
     \label{eq:defEvolOrb}
\end{equation}
or, equivalently, in the matrix form
\begin{equation}
     \mat{C}(t) = \mat{U}(t,t') \mat{C}(t').
     \label{eq:defEvolCmat}
\end{equation}
We can also use the same evolution operator to obtain the solution of the LvN equation in the language of the RDM as
\begin{equation}
     \mat{D}(t) = \mat{U}(t,t')\mat{D}(t')\dg{\mat{U}}(t,t')
     .
     \label{eq:defEvolDmat}
\end{equation}
It is required that the evolution operator is unitary, \emph{i.e.} $\dg{\mat{U}}(t,t')\mat{U}(t,t') = \mathbb{I}$, so that the time evolution preserves the norm of the $\varphi_i$ as well as the idempotence and trace of the density matrix. It follows from the definition that
\begin{align}
        \mat{U}(t,t) &= \mathbb{I}, \label{eq:evolOpProp1}\\
        \mat{U}(t_3,t_1) &= \mat{U}(t_3,t_2)\mat{U}(t_2,t_1), \label{eq:evolOpProp2}\\
        \mat{U}^{-1}(t_1,t_2) &= \mat{U}(t_2,t_1). \label{eq:evolOpProp3}
\end{align}
The last property is related to the TRS described in Section~\ref{sec:TRS} and only holds when no external magnetic fields are present.

Inserting the definition of $\mathbf{U}$ in Eq.~\eqref{eq:defEvolDmat} into the LvN equation recasts the problem of time propagation into determining $\mathbf{U}$ by solving
\begin{equation}
     i\hbar\pd{}{t}\mathbf{U}(t,t') = \mathbf{F}(t)\mathbf{U}(t,t').
     \label{eq:eomEvolOp}
\end{equation}
It is possible to write a closed-form solution of this equation in the form of the Dyson series as
\begin{equation}
        \mathbf{U}(t,t') 
        = 
        \sum_{n=0}^{\infty} \frac{(-i/\hbar)^n}{n!} \int_{t'}^t dt_1\ldots\int_{t'}^t dt_n \mathcal{T}\big\{\mathbf{F}(t_1)\ldots \mathbf{F}(t_n)\big\},
\label{eq:evolutionOpDysonLong}
\end{equation}
where $\mathcal{T}$ represents the time-ordering of the product such, that the leftmost term has the latest time, and each following term is applied at an earlier time than the one before it. The time-ordering is necessary since $\left[\mathbf{F}(t_1),\mathbf{F}(t_2)\right] \neq 0$. Except for the time-ordering, this series represents the expansion of the exponential function and thus is often written in the short-hand form
\begin{equation}
        \mathbf{U}(t,t') = \mathcal{T} \exp \left[-\frac{i}{\hbar}\int_{t'}^t \mathbf{F}(\tau)d\tau \right].
\label{eq:evolutionOpDysonShort}
\end{equation}
This expression, albeit formally exact, requires truncation of the series in numerical implementations. Such truncation inevitably leads to the loss of the unitary property of $\mathbf{U}$, and consequently the idempotence and trace of the density matrix, which can result in numerically unstable time propagation~\cite{castro2004propagators,hochbruck2003magnus,liu2010theoretical}.

\subsection{Magnus expansion}

As an alternative to the Dyson expansion, the evolution operator can be written as a true exponential function that does not require the time ordering, \emph{i.e.} in the form of the exponent of the infinite series as
\begin{equation}
        \mathbf{U}(t,t') = e^{\mathbf{A}(t,t')},
\label{eq:evolutionOpMagnus1}
\end{equation}
where
\begin{equation}
        \mathbf{A}(t,t') = \sum_{n=1}^{\infty}\mathbf{A}_n(t,t').
\label{eq:evolutionOpMagnus2}
\end{equation}
This form was proposed by Magnus in 1954 with the first terms given by~\cite{magnus1954exponential,Tannor2007}
\begin{align}
        \mathbf{A}_1(t,t') 
        =& 
        \frac{1}{i\hbar}\int_{t'}^t dt_1 \mathbf{F}(t_1), 
        \label{eq:magnusA1}\\
        \mathbf{A}_2(t,t') 
        =& 
        -\frac{1}{2}\left(\frac{1}{i\hbar}\right)^2 \int_{t'}^t dt_2 \int_{t'}^{t_2} dt_1 
        \big[\mathbf{F}(t_1),\mathbf{F}(t_2)\big], 
        \label{eq:magnusA2}\\
        \mathbf{A}_3(t,t') 
        =& 
        -\frac{1}{6}\left(\frac{1}{i\hbar}\right)^3 \int_{t'}^t dt_3 \int_{t'}^{t_3} dt_2\int_{t'}^{t_2} dt_1\left( 
        \Big[\mathbf{F}(t_1),\big[\mathbf{F}(t_2),\mathbf{F}(t_3)\big]\Big]
        \right.
        \notag\\
        &+ 
        \left.\Big[\big[\mathbf{F}(t_1),\mathbf{F}(t_2)\big],\mathbf{F}(t_3)\Big]\right).
\label{eq:magnusA3}
\end{align}
Both Dyson and Magnus expansions are equivalent if their respective series are considered in the infinite limit, however, the Magnus expansion has an advantage in the numerical implementations as it retains the unitary property of the evolution operator even if truncated after any number of terms. Higher-order propagators based on the Magnus expansion require the evaluation of multiple commutators of the Hamiltonian (Fock) matrix, \emph{e.g.} in Eqs.~\eqref{eq:magnusA2} and~\eqref{eq:magnusA3}. A commutator-free version based on the Magnus series can be obtained by writing the evolution operator as an (infinite) product of exponentials~\cite{blanes2006fourth,alvermann2011high,gomez2020propagators} which leads to a powerful approach for deriving higher order commutator-free exponential time propagators.

\subsection{Approximate evolution}

Real-time simulations typically start from an initial state defined at $t=0$ by $\mathbf{D}(t\!=\!0)$. This initial state is in most cases obtained from a converged ground-state optimization procedure, though starting the evolution from approximate excited states is also possible~\cite{chapman2011ultrafast}. Once the evolution operator is determined, the orbitals and RDM at arbitrary times can be calculated by applying Eqs.~\eqref{eq:defEvolOrb} and~\eqref{eq:defEvolDmat}. For instance, $\mathbf{D}(t)$ is obtained as follows
\begin{equation}
        \mathbf{D}(t) = \mathbf{U}(t,0)\mathbf{D}(0)\dg{\mathbf{U}}(t,0).
\label{eq:evolDmat}
\end{equation}
In practice, however, the global propagator $\mathbf{U}(t,0)$ is not known -- this is the case even if the infinite series (Dyson or Magnus) from the previous section are truncated. Numerical implementations require that the time is discretized into a finite number of time steps $N$ of size $\Delta t$, and the evolution operator is factored using Eq.~\eqref{eq:evolOpProp2} as
\begin{equation}
    \mathbf{U}(N\Delta t,0) = \prod_{i=0}^{N-1} \mathbf{U}\big((i+1)\Delta t, i\Delta t\big).
\label{eq:evolOpAsProduct}
\end{equation}
Thus, the propagation over one time step is achieved by the application of $\mathbf{U}(t+\Delta t,t)$ on the density matrix or orbitals, where $t\equiv i\Delta t$. This short-time propagation allows us to approximate the integrals in $\mathbf{U}(t+\Delta t,t)$

The most commonly used approximation for $\mat{U}$ is the \emph{midpoint Magnus} propagator~\cite{castro2004propagators,li2005time,wang2007time,kanungo2019real}
\begin{equation}
\mathbf{U}(t+\Delta t,t)
\approx
\exp \left[ \frac{1}{i\hbar} \mathbf{F} \left(t+\frac{\Delta t}{2} \right) \Delta t \right]
,
\label{eq:magnusMidpoint}
\end{equation}
where only the first term $\mat{A}_1$ in Eq.~\eqref{eq:magnusA1} of the Magnus expansion is considered, and the integral $\int_t^{t+\Delta t}dt$ is approximated using the midpoint quadrature. This integrator is of the second order as it is correct to $\mathcal{O}(\Delta t^2)$. Eq.~\eqref{eq:magnusMidpoint} is also the basis of the modified midpoint unitary transformation (MMUT) method in the literature~\cite{li2005time,liang2011efficient}, where the time step is modified to $2\Delta t$. A fourth-order Magnus propagator can be constructed by taking the first two terms in the Magnus series $\vec{A}_1$ and $\vec{A}_2$ and approximating the integrals using a two-point Gaussian quadrature~\cite{castro2004propagators,Ye2022self}. A comparison of the performance of the second- and fourth-order Magnus propagators with the predictor--corrector scheme was presented in the recent study of M{\"u}ller, Sharma, and Sierka
~\cite{muller2020real} based on their efficient implementation of RT-TDDFT. The matrix exponential in Eq.~\eqref{eq:magnusMidpoint} can be evaluated directly by diagonalizing the Fock matrix and constructing the exponent from its eigenvalues. However, techniques that circumvent the expensive diagonalization step by using a series of matrix multiplications, such as the Baker-Campbell-Hausdorff formula~\cite{lopata2011modeling} or the Chebyshev expansion~\cite{williams2016accelerating}, lower the computational cost as well as improve parallelization.

Integrators that are not based on the Magnus series, such as the Runge-Kutta Method~\cite{baer2005time,wang2007time} or the Crank--Nicholson propagator~\cite{crank1947practical,takimoto2007real,schelter2018accurate,lian2018momentum,pemmaraju2020simulation}
\begin{equation}
\mathbf{U}(t+\Delta t,t)
\approx
\frac{1 + i \mathbf{F} \left(t+\frac{\Delta t}{2} \right) \frac{\Delta t}{2\hbar}}{1 - i \mathbf{F} \left(t+\frac{\Delta t}{2} \right) \frac{\Delta t}{2\hbar}}
,
\end{equation}
can also be used. However, unlike the midpoint Magnus or the Crank--Nicholson, the Runge-Kutta integrator does not preserve the unitary property of $\mat{U}$ and can lead to instabilities in the time evolution, since neither the electron number nor the total energy are strictly conserved during a time propagation that is not unitary~\cite{castro2004propagators,goings2018real}. On another note, the exponential Runge-Kutta method~\cite{hochbruck2010exponential} was successfully used to solve equations of motion in the time-dependent coupled-cluster theory~\cite{sato2018communication} which is based on the exponential coupled-cluster parametrization of the wave function~\cite{pedersen2019symplectic,koulias2019relativistic}.

The issue of stable time propagation requires even more attention in the TDHF and TDKS theories. Due to the mean-field, XC, and HF exchange terms, the Fock matrix in the LvN equation depends on the density matrix $\mat{D}(t)$ that is not known at the time $t$. Hence, the LvN equation with this Fock matrix is nonlinear and must be solved self-consistently, \emph{i.e.} the Fock matrix is constructed from the density matrix from the previous iterations (referred to as microiterations in the context of time domain methodology). Unless the loop over microiterations is introduced, this implicit time-dependence of the Fock matrix on the unknown density matrix results in the inability to express the future time midpoint Fock matrix $\mathbf{F} \left(t+\frac{\Delta t}{2} \right)$ that appears in most approximations for the evolution operator. This issue is commonly mitigated by using predictor--corrector or extrapolation--interpolation schemes~\cite{cheng2006simulating,lopata2011modeling,repisky2015excitation,schelter2018accurate,pemmaraju2020simulation}, where the unknown midpoint Fock matrix is first constructed from the previous time step using the linear extrapolation
\begin{equation}
    \mat{F}\left(t+\frac{\Delta t}{2}\right) = 2\mat{F}(t) - \mat{F}\left(t-\frac{\Delta t}{2}\right).
\end{equation}
Once the $\mat{D}(t+\Delta t)$ is obtained by applying the evolution operator $\mat{U}(t+\Delta t,t)$, a new Fock matrix at $t+\Delta t$ is formed. The two Fock matrices at $t$ and $t+\Delta t$ are then linearly interpolated to create the updated midpoint Fock matrix
\begin{equation}
    \mat{F}\left(t+\frac{\Delta t}{2}\right) = \frac{1}{2}\mat{F}(t) + \frac{1}{2}\mat{F}\left(t+\Delta t\right),
\end{equation}
and the time propagation restarts from the initial time $t$. This process is repeated until the self-consistence is reached. A thorough comparison of various propagation schemes in the context of nonrelativistic TDKS can be found in the study of Pueyo \emph{et al.}~\cite{gomez2018propagators}. 
A detailed analysis is provided in the chapter by Ye \emph{et al.}~\cite{Ye2022self}.
Simulations of X-ray absorption spectra are particularly sensitive to the size of the time step as the core excitations appearing in the high energy region typically involve rapid oscillations of the wave functions that require a very small time step to be described properly~\cite{Kadek2015,Kasper2018Modeling,yang2022intruder,Ye2022self}. For such studies, Ye \emph{et al.}~\cite{Ye2022self} presented a relativistic X2C approach based on the fourth-order commutator-free Magnus propagator that adaptively chooses optimal time step and simulation time. The discussion of the approximate time evolution is expanded further in this volume in the chapter by L.~Ye, H.~Wang, Y.~Zhang, Y.~Xiao and W.~Liu entitled ``Real-Time Time-Dependent Density Functional Theories with Large Time Step and Short Simulation Time''.


\section{Signal processing}
\label{sec:SignalProcessing}

RT-TDSCF solves the EOM by direct propagation in the time domain.
This allows direct simulation of time-resolved experiments and to obtain the entire spectral information from a single real-time calculation.
On the other hand, physical quantities of experimental interest are often defined
in the frequency domain, which creates a demand for techniques that efficiently extract the frequency-domain quantity from the simulations that are carried out with finite numerical accuracy, time step length, and simulation time.

A time-dependent property $f(t)$ can be translated to the frequency domain as the Fourier integral
\begin{equation}
    \tilde{f}(\omega) = \int^{\infty}_{-\infty} f(t) e^{i\omega t} dt.
\label{eq:td2fdfourier}
\end{equation}
However, due to the presence of periodic oscillations in $f(t)$ that originate in the quantum mechanical evolution containing the excitation energies, the integral in Eq.~\eqref{eq:td2fdfourier} leads to $\delta$-functions in the frequency domain. Such stick spectra are difficult to describe in numerical simulations, hence, the Fourier integral is replaced with the Laplace transform
\begin{equation}
    \tilde{f}(\omega) = \int^{\infty}_0 f(t) e^{i\omega t-\gamma t} dt,
\label{eq:td2fdlaplace}
\end{equation}
where we assumed that there is no response of the studied system before the perturbation is applied at time $t=0$, and we introduced a phenomenological damping parameter $\gamma>0$. This damping parameter accounts for the fact that practical real-time simulations are performed with a finite simulation time, and truncating the periodically oscillating signal that is not damped results in undesirable features in the spectrum. To this end, $\gamma$ is empirically set to a value inversely proportional to the simulation time to ensure that the oscillations diminish before the simulation is terminated. Analytic evaluation of Eq.~\eqref{eq:td2fdlaplace} for periodic signals leads to a series of Lorentzian peaks with finite width in the spectrum.

Signal processing techniques serve as a means to extract the frequency-dependent molecular properties $\tilde{f}(\omega)$ from the simulation results $f(t)$, \emph{i.e.} to approximate the integrals in Eqs.~\eqref{eq:td2fdfourier} and~\eqref{eq:td2fdlaplace}.
In this section we present an overview of approaches used in the context
of RT-TDSCF, starting with the simplest one, the Discrete Fourier transform, that is
instructive in order to explain the relationship between the time and frequency domains,
and then moving to the more sophisticated methods that achieve better resolution
from a shorter simulation.
In general, spectra with a high density of states pose a bigger challenge to
the signal processing methods.

\subsection{Discrete Fourier transform}

In practice, RT-TDSCF simulations are performed in a series of discrete time steps $t_j$ for which the induced dipole moment is calculated from a trace of the
dipole moment matrix and the time-dependent density matrix
\begin{equation}
\bm{\mu}^\mathrm{ind}(t_j) = \Tr[\mathbf{P}\mathbf{D}(t_j)] - \bm{\mu}^\mathrm{static}
,
\end{equation}
where the static dipole moment is calculated as $\bm{\mu}^\mathrm{static} = \Tr[\mathbf{P}\mathbf{D}_0]$.
The most straightforward approximation to Eq.~\eqref{eq:td2fdlaplace} is the discrete Fourier transform
\begin{equation}
\label{eq:freqTransfDef}
\tilde{f}_k
=
\sum_{j=0}^{n-1} \Delta t\, f_j e^{2\pi i \frac{jk}{n} - \gamma j\Delta t}
.
\end{equation}
Here, $k = 0, 1, \ldots, n-1$ where $n$ is the number of time steps and $\omega_k = 2\pi k / (n\Delta t)$ 
is the $k$-th frequency point. The coefficients $f_j \equiv \bm{\mu}^\mathrm{ind}(t_j)$ and $\tilde{f}_k \equiv \bm{\mu}^\mathrm{ind}(\omega_k)$ represent the components of the induced dipole moment in time and frequency domains, respectively.

\subsection{Relationship between the time and frequency domains}

The frequency-domain results are obtained by discrete Fourier transform of the time-domain
results. If we perform a time-domain simulation that consists of $n$ steps of length
$\Delta t$, the Fourier transform yields a frequency-domain interval of length
\begin{equation}
\label{eq:freqInterval}
\Omega
=
\frac{2\pi}{\Delta t}
.
\end{equation}
Since the number of points in both domains is the same, the resolution in the frequency
domain is
\begin{equation}
\Delta\omega
=
\frac{2\pi}{n \Delta t}
.
\label{eq:resolution}
\end{equation}
This relationship tells us that in order to increase the resolution in calculated spectra
we need to increase the total simulation length $n\Delta t$ by increasing the number of
time steps (which makes the simulation more time consuming) or increasing the size of the
time step (which puts extra demands on the solver). However, because the frequency-domain
interval depends inversely on the time-step length, see Eq.~\eqref{eq:freqInterval},
in order to describe high-frequency parts of spectra, such as in X-ray spectroscopies,
shorter time steps are required. Therefore, a balance between the resolution, frequency
range and computational cost has to be achieved by choosing suitable simulation parameters.
Eq.~\eqref{eq:resolution} represents the major limitation for obtaining spectra with high resolution when using the discrete Fourier transform.

\subsection{Pad\'{e} approximants}

The use of the Pad\'{e} approximation as a signal processing technique~\cite{dey1998efficient,guo2001computation} was introduced to RT-TDSCF
by Bruner, LaMaster, and Lopata~\cite{Bruner2016}
and quickly gained popularity due to its advantage compared to the widespread Fourier transform.
In the Pad\'{e} approximants, the expression for the Fourier components $\tilde{f}(\omega)$ (\emph{e.g.} induced dipole moment in frequency domain)
\begin{equation}
    \tilde{f}(\omega) = \sum_{j=0}^{M} f(t_j) \Delta t e^{i\omega j\Delta t} e^{-\gamma j\Delta t}
    ,
\end{equation}
is understood as a power series
\begin{equation}
    \label{eq:PadeSeries}
    \tilde{f}(\omega) = \sum_{j=0}^{M} c_j (z_\omega)^j
    ,
\end{equation}
where $z_\omega = e^{i\omega \Delta t}$ and $c_j = f(t_j) \Delta t e^{-\gamma j\Delta t}$
that can be approximated as a division of two other power series using the Pad\'{e} ansatz
\begin{equation}
    \label{eq:PadeAnsatz}
    \tilde{f}(z) = \frac{\sum_{k=0}^{N} a_k z^k}{\sum_{k'=0}^{N} b_{k'} z^{k'}}
    ,
\end{equation}
where $N=M/2$.
The comparison of Eqs.~\eqref{eq:PadeSeries} and \eqref{eq:PadeAnsatz} leads to a system of
equations, or a matrix equation, for the coefficients $b$
\begin{equation}
    \mat{b} = \mat{G}^{-1} \mat{d}
    ,
\end{equation}
where $G_{km} = c_{N-m+k}$ and $d_k = -c_{N+k}$.
The system is overdetermined, leading to a customary choice of setting $b_0 = 1$.
The knowledge of the $b$-coefficients can then be used to determine the $a$-coefficients
from $a_k = \sum_{m=0}^k b_m c_{k-m}$. The $a$ and $b$ coefficients are subsequently
used to approximate the Fourier transform $\tilde{f}(\omega)$ with the advantage that
$a$ and $b$ are not functions of frequency. Hence, the frequency can be chosen
at will without the limitation of Eq.~\eqref{eq:resolution}, which allows
the spectrum to be evaluated with arbitrary frequency resolution.
In practice, the Pad\'{e} approximation can suffer from numerical instabilities.
To mitigate this, the original work~\cite{Bruner2016} suggested to combine it with a MO-based decomposition~\cite{repisky2015excitation},
where each occupied--virtual MO pair that contributes to the net frequency-dependent polarizability
is transformed into the frequency domain using its own Pad\'{e} approximation, \emph{i.e.}
the coefficients $a_k$ and $b_k$ become different for each MO pair.
A spectrum constructed from an individual MO pair is typically sparser and thus
less prone to defective behaviour.
The final spectrum is then a sum of spectra over all MO pairs.

\subsection{Compressed sensing}

Compressed sensing is a technique based on the observation that a small number of points in time domain is sufficient to sample a frequency domain signal that is sparse, \emph{i.e.} when many Fourier-domain coefficients $\tilde{f}_k$ in Eq.~\eqref{eq:freqTransfDef} are near zero. The application of compressed sensing in RT-TDDFT was first explored by Andrade, Sanders, and Aspuru-Guzik~\cite{andrade2012application} in the context of electronic and nuclear dynamics for the calculation of vibrational and optical spectra. The compressed sensing method recasts the problem of finding the Fourier coefficients into solving a system of linear equations
\begin{equation}
    \mat{A}\tilde{\vec{f}} = \vec{f},
\end{equation}
where $f_j \equiv (\vec{f})_j$ and $\tilde{f}_k \equiv (\tilde{\vec{f}})_k$ are vectors with components in time and frequency domains, respectively, and $\mat{A}$ is the matrix containing the complex phase factors. This system allows for a different number of time and frequency points $j=0,1,\ldots,n_t-1$ and $k=0,1,\ldots,n_{\omega}-1$. For a small number of time points $n_t < n_{\omega}$ the system is underdetermined with infinitely many solutions $\tilde{\vec{f}}$. The sparse solution with the largest number of zero coefficients is then obtained by finding $\tilde{\vec{f}}$ that minimize the norm $|\tilde{f}|$ while satisfying
\begin{equation}
    \left|\mat{A}\tilde{\vec{f}} - \vec{f}\right| < \eta,
\end{equation}
where $\eta \ll 1$ accounts for a certain amount of numerical noise in the signal. Even though the benefits of compressed sensing are expected to be lower for dense signals, for systems with low density-of-states, significant savings can be obtained by reconstructing the spectra from shorter time simulations~\cite{andrade2012application}.

\subsection{Filter diagonalization}

Time signals often take the form of a sum of damped oscillations, \emph{i.e.}
\begin{equation}
    f(t) = \sum_m d_m e^{-i\omega_m t},
\label{eq:harmonicInv}
\end{equation}
where the frequencies $\omega_m$ can be considered complex to account for the damping factors. In an ideal case (with no numerical noise), this is also the form obtained from the quantum mechanical time propagation discussed Section~\ref{sec:propagation}, with $\omega_m$ and $d_m$ corresponding to excitation energies and oscillator strengths, respectively. Determining the values of $\omega_m$ and $d_m$ from the known signal $f(t)$ is referred to as the \emph{harmonic inversion} problem~\cite{mandelshtam1997harmonic}, and it was the connection to the quantum mechanical evolution that lead to the formulation of harmonic inversion as an eigenvalue problem~\cite{wall1995extraction,mandelshtam1997low,mandelshtam2001fdm}. The signal $f(t)$ in Eq.~\eqref{eq:harmonicInv} can be considered a correlation function
\begin{equation}
    f(t) = \braket{\psi_0|e^{-i\hat{\Omega} t}|\psi_0},
\end{equation}
some unknown Hamiltonian $\hat{\Omega}$ and an initial state $\psi_0$. The frequencies $\omega_m$ are the eigenvalues of $\hat{\Omega}$ and are obtained by solving the generalized eigenvalue equation
\begin{equation}
    \mat{U}\vec{b}_m = u_m \mat{S}\vec{b}_m,
\label{eq:krylovEval}
\end{equation}
where $\hat{U} = e^{-i\hat{\Omega} \Delta t}$, $u_m = e^{-i\omega_m\Delta t}$, and $\Delta t$ is the time step. This equation can be formulated in Krylov basis constructed by a consecutive application of the operator $\hat{U}$ on a reference vector $\vec{v}_0$ as $\vec{v}_j := \hat{U}^j \vec{v}_0$. In such a basis, the matrices $\mat{U}$ and $\mat{S}$ take simple forms of $U_{jj'} = f_{j+j'+1}$ and $S_{jj'} = f_{j+j'}$, respectively, obtained from the time signal as $f_j \equiv f(j\Delta t)$. The coefficients $d_m$ are calculated using the eigenvectors $\vec{b}_m$ as $\sqrt{d_m} = \vec{b}_m^T\vec{f}$. Note, that this method assumes that the time signal has the form of Eq.~\eqref{eq:harmonicInv}, but it is not necessary that the signal was generated by an actual propagation with a quantum mechanical Hamiltonian, nor do we need to know the explicit forms of $\hat{\Omega}$ or $\psi_0$.

Even though Eq.~\eqref{eq:krylovEval} provides a formally exact solution to the harmonic inversion problem (up to the dimension of the Krylov vector space), it suffers from the cubic scaling $\mathcal{O}(n^3)$ of the diagonalization procedure with the number of time steps compared to $\mathcal{O}(n\log n)$ scaling of the Discrete Fourier Transform. The filter diagonalization method circumvents this problem by transforming the matrices in Eq.~\eqref{eq:krylovEval} from the Krylov basis $\vec{v}_j$ into the Fourier basis
\begin{equation}
    \vec{w}_k = \sum_{j=0}^{n-1} e^{ij\Delta t \xi_k} \vec{v}_j,
\label{eq:krylovToFourier}
\end{equation}
where the frequencies $\xi_k$ can be chosen to form an equidistant grid. Since the basis vectors $\vec{w}_k$ are localized in the frequency domain, the eigenvectors of the operator $\hat{U}$ can be expressed using a small number $k=1,\ldots, n_\text{win} \ll n$ of $\vec{w}_k$. Nonnegligible contributions to the $m$-th eigenvector arise only from basis vectors for which $\xi_k\approx\omega_m$. Thus, matrices $\mat{U}$ and $\mat{S}$ expressed in this basis exhibit large diagonal and diminishing off-diagonal terms. This enables defining a small spectral window $\left[\omega_\text{min},\omega_\text{max}\right]$ of $n_\text{win}$ frequencies  and diagonalizing the matrix $\mat{U}$ in Eq.~\eqref{eq:krylovEval} in the Fourier subspace spanned by $\vec{w}_k$. The main disadvantage of the filter diagonalization method is the assumption that the time signal takes the form of Eq.~\eqref{eq:harmonicInv}. Even though the method can efficiently circumvent the uncertainty relation in Eq.~\eqref{eq:resolution} and provide high resolution spectra of sparsely distributed dominant peaks, practical real-time simulations are hampered by numerical noise, which complicates the use of the filter diagonalization method for extracting spectral information in highly dense regions.

\section{Calculation of molecular properties using real-time methods}
\label{sec:applications}


In this section we review some of the areas where relativistic RT-TDSCF methods have been
employed to combine the advantages of both the relativistic and real-time treatments.
Moreover, our aim is to explain the physical context of the molecular properties,
to show how to construct a computational protocol for obtaining these properties,
and what aspects of the calculations to pay attention to.
First we explore the calculation of linear and non-linear response properties where real-time
propagation is an alternative to perturbation theory.
Then we focus on non-equlibrium spectroscopies, for which real-time methods are the only viable approach.
We focus on heavy-element systems and X-ray spectroscopies, where relativistic effects are paramount.
Such applications are the primary motivation behind the development of relativistic real-time methods.

\subsection{Linear response properties}

Linear response properties include some of the most commonly measured spectroscopies
such as electron absorption spectroscopy (EAS), including X-ray absorption spectroscopy (XAS),
and chiroptical spectroscopies such as electron circular dichroism (ECD) and optical
rotatory dispersion (ORD).~\cite{Norman2018}
Therefore, they are both experimentally relevant as well as
provide a good introduction to the workflow and analysis of real-time simulations.

In this section, we consider a molecular system perturbed by a single external field
$\vec{\mathcal{E}}(t)$
interacting with the molecule within a dipole approximation which induces a time-dependent
dipole response in the molecule.
The response is a physical quantity $R(t)$ calculated as an expectation value of its operator
$\hat{R}$ from the time-dependent wave function, $R(t) = \braket{\Psi(t)|\hat{R}|\Psi(t)}$.
In cases when $\hat{R}$ is a one-electron operator, such as the electric dipole operator, its expectation
value can be calculated as the trace that contains the time-dependent one-electron RDM
$R(t) = \Tr[\mat{D}(t)\mat{R}]$.
Restricting ourselves to electric and magnetic dipole moment operators for the perturbation and response,
different combinations lead to different linear spectroscopies.
First, we follow the induced electric dipole resulting from an electric dipole perturbation,
leading to EAS spectrum and the frequency-dependent index of refraction.

\paragraph{Electron absorption spectroscopy}
Electron absorption spectrum at all frequencies from UV/Vis to X-ray is determined by
the complex frequency-dependent polarizability tensor $\bm{\alpha}(\omega)$.
The polarizability tensor connects the induced electric dipole moment to
an applied electric field, which in the frequency domain reads
\begin{equation}
\label{eq:FDPolDef}
\mu^\mathrm{ind}_u (\omega)
=
\alpha_{uv} (\omega) \mathcal{E}_v (\omega) + \ldots
,
\end{equation}
where $\mathcal{E}_v (\omega)$ is the Fourier transform of the external electric field
$\vec{\mathcal{E}}(t) = \mathcal{E} \vec{n} F(t)$ defined by its amplitude $\mathcal{E}$,
directional unit vector $\vec{n}$, and time dependence $F(t)$.
The external field couples to the molecular system via the electric dipole operator, resulting
in its appearance in the Fock matrix as the term
\begin{equation}
    V^\mathrm{ext}(t)
    =
    -\mathcal{E} F(t) \vec{n} \cdot \mat{P}
\end{equation}
where $\mat{P}$ is a matrix representation of the electric dipole moment operator.
By connecting the molecular induced dipole moment, i.e. the polarization of a bulk material,
to the applied external field, the polarizability tensor determines the complex index of refraction
whose real part is the standard index of refraction while the imaginary part corresponds to the
attenuation coefficient describing the absorption of light and appearing in the Beer--Lambert law.
However, the more common way of expressing the absorption spectrum is via the photoabsorption
cross-section tensor
\begin{equation}
\bm{\sigma}(\omega)
=
\frac{4\pi\omega}{c} \Im \left[ \bm{\alpha}(\omega) \right]
,
\end{equation}
where $\Im$ denotes the imaginary part, and $c$ is the speed of light.
The absorption spectrum is then the dipole strength function obtained from the rotational average of the tensor $\bm{\sigma}$
\begin{equation}
\label{eq:DipoleStrengthFun}
S(\omega)
=
\frac{1}{3} \left[ \Tr \bm{\sigma}(\omega) \right]
,
\end{equation}
where Tr is the trace over the Cartesian components.

\begin{figure}
    \centering
    \includegraphics[width=0.95\textwidth]{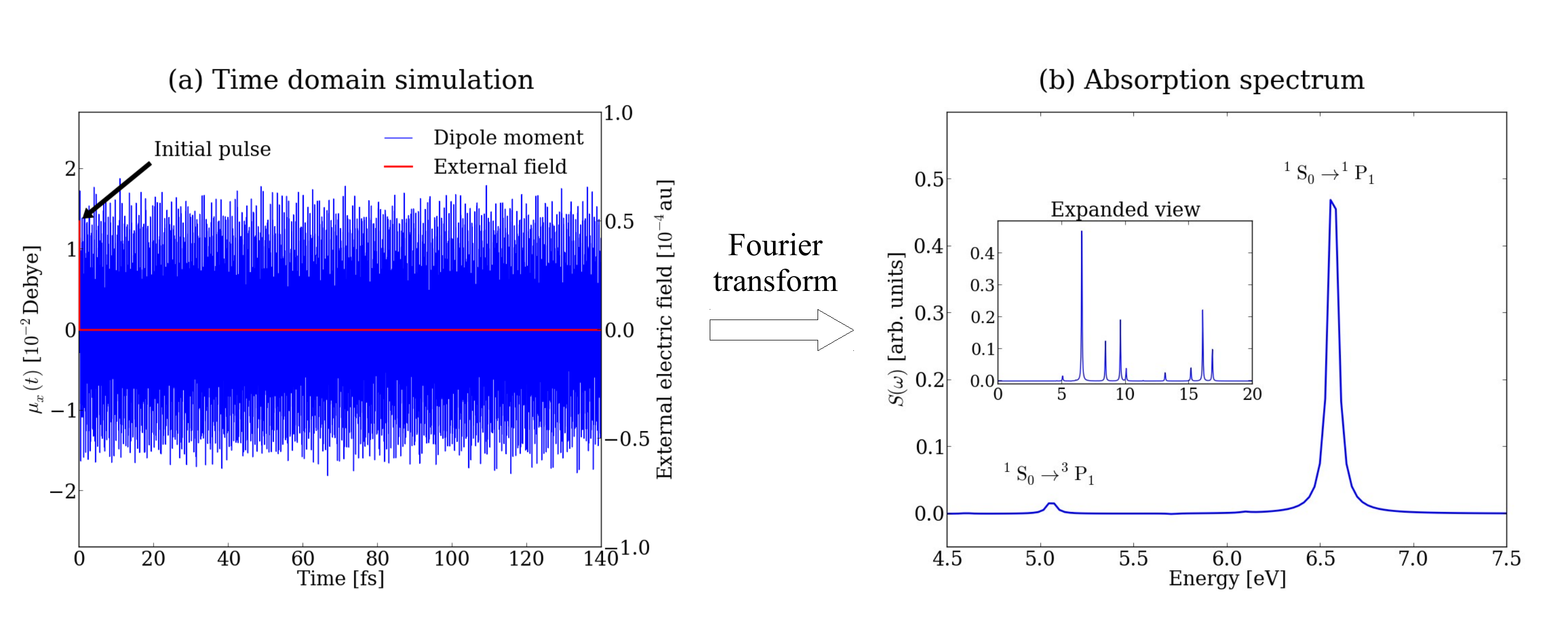}
    \caption{Calculated absorption spectrum of the Hg atom using the relativistic four-component RT-TDDFT.
             The electronic ground state is perturbed by a $\delta$-type pulse, the induced dipole moment is recorded
             in time and transformed to the frequency domain to yield the spectrum. A formally forbidden
             singlet--triplet transition is present due to the relativistic level of theory.
             Reprinted with permission from Ref.~[\citenum{repisky2015excitation}]. Copyright 2015, American Chemical Society.}
    \label{fig:SpectrumWorkflow}
\end{figure}

A calculation of the absorption spectrum defined in Eq.~\ref{eq:DipoleStrengthFun} from
a real-time simulation then proceeds in the following steps:
\begin{enumerate}
    \item Obtain the reference ground-state density matrix $\mat{D}_0$ by solving the time-independent SCF equation.
    \item Perturb the ground state to obtain the initial state $\mat{D}(t_0)$.
    This is usually performed by a short ``kick'' in the time domain
    that corresponds to a broadband pulse in the frequency domain, thus exciting all molecular transitions.
    A pure form of such a pulse is the Dirac $\delta$ function
    $\vec{\mathcal{E}}(t) = \mathcal{E} \vec{n} \delta(t-t_0)$
    which in practical simulations can be represented numerically by a narrow
    Gaussian function or rectangle, or by an analytic expression 
    \begin{equation}
        \mat{D}(t_0) = e^{i\mathcal{E}\vec{n}\cdot\mat{P}/\hbar} \mat{D}_0 e^{-i\mathcal{E}\vec{n}\cdot\mat{P}/\hbar}
        .
    \end{equation}
    which represents an infinitesimally short time evolution by $\mat{U}(t_0+\varepsilon,t_0-\varepsilon)$ driven by the $\delta(t-t_0)$ field in the limit $\varepsilon\rightarrow 0$~\cite{repisky2015excitation}.    
    \item Propagate the density matrix $\mat{D}(t_0)$ in time for $n$ time steps of length $\Delta t$
    while recording the induced dipole moment $\vec{\mu}(t) = \Tr[\mat{D}(t)\mat{P}]$ at each time step.
    \item Transform the induced dipole moment to the frequency domain using some of the techniques
    discussed in Section~\ref{sec:SignalProcessing}, i.e. calculate
    \begin{equation}
       \vec{\mu}(\omega) = \int_{t_0}^{\infty} dt \vec{\mu}(t) e^{i\omega t - \gamma t},
    \end{equation}
    where the damping term $e^{-\gamma t}$ is introduced to resolve the problem that arises when
    periodic signals are truncated in numerical simulations with finite time length.
\end{enumerate}
A graphical summary of these steps is shown in Figure~\ref{fig:SpectrumWorkflow} for the case
of the EAS spectrum of the mercury atom (SVWN5 functional~\cite{slater1951, vosko1980}, uncontracted Dyall's VDZ basis~\cite{dyall2010-5d})
calculated from a four-component RT-TDDFT simulation.
Besides illustrating the workflow of an EAS calculation from RT-TDDFT, the figure also demonstrates
the importance of including relativistic effects in such simulations by capturing the formally
forbidden singlet--triplet transition.
Even though non-relativistic approaches to these transitions have been presented employing for example
a spin-dependent perturbation,~\cite{isborn2009singlet}
it is only in relativistic theories that include spin--orbit coupling that singlet--triplet transitions appear
in the spectra naturally from first principles and with correct intensities.
Therefore, several works on relativistic RT-TDSCF for electron absorption spectroscopy have focused on describing
singlet--triplet transitions in the spectra.~\cite{repisky2015excitation, Goings2016, DeSantis2020pyberthart}

\paragraph{X-ray absorption}
X-ray absorption spectroscopy (XAS) is a subset of electron absorption spectroscopy where high-frequency X-ray radiation is absorbed in molecules
while exciting electrons from core orbitals. Therefore, in XAS the same physical quantities are evaluated.
However, relativistic effects, both scalar relativistic effects manifesting as shifts of spectral lines, as well as spin--orbit
interaction causing the splitting of spectral lines, are more pronounced in XAS necessitating the use of computational methods based
on relativistic Hamiltonians. These effects are observable even in light (3rd row) elements~\cite{Kadek2015}, highlighting the need for
a relativistic description also in these cases.
The computational protocol for calculating XAS is the same as presented in the previous paragraph for EAS in the UV/Vis frequency range with two important caveats:
(i) X-ray absorption occurs at higher frequencies so that the settings of the simulation such as time step and
the number of time steps have to be adjusted in order to reach the desired frequencies with sufficient resolution
and numerical accuracy;
(ii) in simulations using finite atom-centered basis sets, a broadband $\delta$-type pulse excites all molecular modes
including excitations from valence orbitals to high-lying above-ionization virtuals that may fall into the XAS frequency range,
but are non-physical relicts of an improper description of continuum states, and thus have to be eliminated either in post-processing
or during the application of the external field.~\cite{Kadek2015,yang2022intruder}

\begin{figure}
    \centering
    \includegraphics[width=0.7\textwidth]{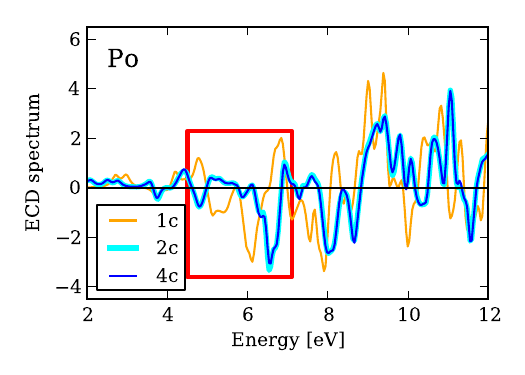}
    \caption{ECD spectrum of polonium analogue of dimethyloxirane calculated using relativistic 4c as well as 2c RT-TDDFT compared to
    a non-relativistc 1c result. The inclusion of relativistic effects changes the sign of the spectrum in the highlighted
    area demonstrating the need of proper treatment of relativity when addressing chiroptical spectroscopic properties.
    Adapted from Ref.~\citenum{Konecny-JCP-149-204104-2018} with permission. Copyright 2018, AIP.}
    \label{fig:DMPoECD}
\end{figure}

\paragraph{Chiroptical spectroscopies}
Optical activity and circular dichroism are effects arising when chiral matter interacts with polarized light.
Chiral molecules possess a different complex index of refraction for right- and left-handed circularly polarized (CP) light.
The real part determines the different refraction of  CP light and also the rotation of the plane of polarization of
linearly polarized (LP) light, while the imaginary part determines the difference in absorption of CP light and the induced ellipticity
of LP light.~\cite{Barron2004}
At the molecular level, the property underpinning these processes is the electric dipole--magnetic dipole tensor $\mat{\beta}$
(also known as Rosenfeld tensor), that also connects to the first order the induced electric dipole moment $\vec{\mu}^\mathrm{ind}$ to the time derivative
of the external magnetic field $\vec{B}$ as well as the induced magnetic dipole moment $\vec{m}^\mathrm{ind}$ to the time derivative of the external electric field $\mat{E}$
\begin{alignat}{2}
\label{eq:elDipoledBdt}
\mu^\mathrm{ind}_i (\omega) & =   && \beta_{ij} (\omega) \dot{B}_j (\omega),
\\
\label{eq:magDipoledEdt}
m^\mathrm{ind}_i (\omega)   & = - && \beta_{ji} (\omega) \dot{E}_j (\omega)
.
\end{alignat}
Note that we have restricted ourselves to isotropic samples where a quadrupolar contribution that is non-zero for a single molecule
vanishes after averaging over molecular orientations.
RT-TDSCF calculations of chiroptical properties are based on Eq.~\eqref{eq:magDipoledEdt} rather than on the direct simulation
of an interaction of molecules with circularly polarized light.
The calculation proceeds analogously to the computational protocol outlined here for electron absorption spectroscopy:
a molecule in its ground state $\mat{D}_0$ is perturbed by an external electric field in the form of a $\delta$-pulse
and the induced \emph{magnetic} dipole moment
\begin{equation}
\label{eq:magDipoleTrace}
\vec{m}^\mathrm{ind}(t) = \Tr[\mat{M}\mat{D}(t)] - \bm{m}^\mathrm{static}
,
\end{equation}
is evaluated in the course of the simulation.
In Eq.~\eqref{eq:magDipoleTrace}, $\vec{m}^\mathrm{static} = \Tr[\mat{M}\mat{D}_0]$ is the static magnetic dipole moment
and
\begin{equation}
   \mat{M}^\mathrm{4c}_{\mu\nu} =
   -\frac{1}{4c}
   \begin{pmatrix}
   \bm{0}       &   \braket{X_\mu | (\vec{r}_\mathrm{g}\times\vec{\sigma}) (\vec{\sigma}\cdot\vec{p}) | X_\nu} \\
   \braket{X_\mu | (\vec{\sigma}\cdot\vec{p}) (\vec{r}_\mathrm{g}\times\vec{\sigma}) | X_\nu}  &  \bm{0}
   \end{pmatrix}
   ,
\end{equation}
is the matrix representation of the magnetic dipole moment operator in the RKB basis with $\vec{r}_\mathrm{g} = \vec{r}-\vec{R}_\mathrm{g}$
standing for the electronic position operator relative to a fixed gauge $\vec{R}_\mathrm{g}$.
The induced magnetic dipole moment is transformed to the frequency domain and used to calculate
the Rosenfeld tensor via
\begin{equation}
\label{eq:betaFromM}
\beta_{ji}(\omega)
=
- i \frac{m^\mathrm{ind}_i (\omega)}{\mathcal{E}}
,
\end{equation}
where $\mathcal{E}$ is again the amplitude of the perturbing external field.
Chiroptical properties are notoriously sensitive to different parameters of a calculations such as the choice of functional,
basis set, conformation of the molecule, solvent effects etc.~\cite{Warnke2012circular}. Using relativistic RT-TDDFT it was shown~\cite{Konecny-JCP-149-204104-2018}
on a series of model molecules -- analogs of dimethyloxirane with the oxygen atom replaced with heavier homologues (S, Se, Te, Po, Lv),
that relativity alone can change the sign of the spectral function, i.e. the factor discriminating between the enantiomers.
An example of such a spectrum is shown in Figure~\ref{fig:DMPoECD} for dimethylpolonirane (PBE functional~\cite{slater1951, Perdew1996, Perdew1997},
uncontracted Dyall's aug-cVDZ basis~\cite{Dyall1998-4-6p, Dyall2006-4-6p-rev} for Po, and uncontracted Dunning's aug-cc-pVDZ~\cite{Dunning1989, Kendall1992} for light elements).
Therefore, relativistic real-time methods should be an important tool in practical calculations of chiroptical spectra,
especially of molecules containing heavy elements.

\subsection{Nonlinear optical properties}

For a weak external field, the spectra resulting from the real-time propagation will be equivalent to the results obtained using response theory.
However, in stronger fields, real-time simulations contain corrections of higher orders.
This is seen from comparing the perturbation expansion for the induced dipole moment,
schematically
\begin{equation}
  \label{eq:IndDipoleOrders}
  \vec{\mu}^\mathrm{ind}
  =
  \bm{\alpha}^\mathrm{PT} \vec{\mathcal{E}} + \bm{\beta}^\mathrm{PT} \vec{\mathcal{E}}^2 + \bm{\gamma}^\mathrm{PT} \vec{\mathcal{E}}^3 + \ldots
,
\end{equation}
with the way the induced dipole moment from a real-time simulation is processed,
again schematically
\begin{equation}
  \label{eq:RTDipoleOrders}
  \vec{\mu}^\mathrm{ind}
  =
  \bm{\alpha}^\mathrm{RT} \vec{\mathcal{E}}
  =
  \left[ \bm{\alpha}^\mathrm{PT} + \bm{\beta}^\mathrm{PT} \vec{\mathcal{E}} + \bm{\gamma}^\mathrm{PT} \vec{\mathcal{E}}^2 + \ldots \right] \vec{\mathcal{E}}
  .
\end{equation}
In Eqs.~\eqref{eq:IndDipoleOrders} and~\eqref{eq:RTDipoleOrders} the indices PT and RT refer to perturbation theory and real-time, respectively,
and the molecular properties correspond to
polarizability ($\bm{\alpha}$), first hyperpolarizability ($\bm{\beta}$), and second hyperpolarizability ($\bm{\gamma}$). 
While this feature of real-time methods enables the study of strong-field effects in spectra,
the properties of higher orders are incorporated in the non-perturbative $\vec{\mu}^\mathrm{ind}$ or $\bm{\alpha}^\mathrm{RT}$
and are not readily available for further analysis.
However, in some applications, it is desirable
to know the values of higher-order responses individually. This is also possible
to achieve using real-time methods by combining simulations with various field strengths.

To show how a method for obtaining nonlinear responses from real-time simulations can work,
let us examine more closely the Taylor expansion of the time-dependent induced dipole moment
\begin{equation}
\label{eq:responses}
\mu_i(t) = \mu_{ij}^{(1)}(t) \mathcal{E}_j + \mu_{ijk}^{(2)}(t) \mathcal{E}_j \mathcal{E}_k + \mu_{ijkl}^{(3)}(t) \mathcal{E}_j \mathcal{E}_k \mathcal{E}_l + \ldots
\end{equation}
where $\mathcal{E}_j$ combines the amplitude and direction of the external field, i.e. $\vec{\mathcal{E}}= \mathcal{E}\vec{n}$,
and we defined the $n$-th order contributions $\mu^{(n)}$ to the induced dipole moment.
These contributions are convolutions of the time-dependent (hyper)polarizability tensors with
the time dependence of the external field(s)
\begin{subequations}
\label{eq:NorderTD}
\begin{flalign}
\label{eq:1orderTD}
\mu_{ij}^{(1)}(t)   & = \int\! dt_1\, \alpha_{ij}(t-t_1) F(t_1),  \\
\label{eq:2orderTD}
\mu_{ijk}^{(2)}(t)  & = \frac{1}{2!} \int\! dt_1 \int\! dt_2\, \beta_{ijk}(t-t_1,t-t_2) F(t_1) F(t_2), \\
\label{eq:3orderTD}
\mu_{ijkl}^{(3)}(t) & = \frac{1}{3!} \int\! dt_1 \int\! dt_2 \int\! dt_3\, \gamma_{ijkl}(t-t_1,t-t_2,t-t_3) F(t_1) F(t_2) F(t_3) .
\end{flalign}
\end{subequations}
Again, the experimentally relevant quantities are the frequency-dependent (hyper)polarizability tensors
\begin{subequations}
\label{eq:propertiesFT}
\begin{flalign}
\alpha_{ij}(\omega)                          = & \int\! dt_1\, \alpha_{ij}(t_1)\, e^{-i \omega t_1}, \\
\beta_{ijk}(\omega_1, \omega_2)              = & \int\! dt_1 \int\! dt_2\, \beta_{ijk}(t_1, t_2)\, e^{-i \omega_1 t_1} e^{-i \omega_2 t_2}, \\ 
\gamma_{ijkl}(\omega_1, \omega_2, \omega_3 ) = & \int\! dt_1 \int\! dt_2 \int\! dt_3\, \gamma_{ijkl}(t_1,t_2,t_3)\, e^{-i \omega_1 t_1} e^{-i \omega_2 t_2} e^{-i \omega_3 t_3}
.
\end{flalign}
\end{subequations}
If we choose a harmonic external field,
$V^\mathrm{ext}(t) = \mathcal{E} \cos(\omega t) \vec{n} \cdot \mat{P}$,
the integrals in Eqs.~\eqref{eq:NorderTD} and \eqref{eq:propertiesFT} can be simplified to obtain expressions
relating $\mu^{(n)}$ to specific nonlinear optical (NLO) properties
\begin{subequations}
\label{eq:NorderFD}
\begin{flalign}
\label{eq:1orderFD}
\mu_{ij}^{(1)}(t)   & = \alpha_{ij}(-\omega;\omega) \cos(\omega t) ,  \\
\label{eq:2orderFD}
\mu_{ijk}^{(2)}(t)  & = \frac{1}{4} \left[ \beta_{ijk}(-2\omega;\omega,\omega) \cos(2\omega t) + \beta_{ijk}(0;\omega,-\omega) \right] , \\
\label{eq:3orderFD}
\mu_{ijkl}^{(3)}(t) & = \frac{1}{24} \left[ \gamma_{ijkl}(-3\omega;\omega,\omega,\omega) \cos(3\omega t)
                           + 3\bar{\gamma}_{ijkl}(-\omega;\omega,\omega,-\omega) \cos(\omega t) \right] .
\end{flalign}
\end{subequations}
The frequency-dependent molecular property tensors in equations \eqref{eq:NorderFD}
are the dipole polarizability $\alpha_{ij}(-\omega;\omega)$,
and higher-order properties governing processes involving several photons,
namely, the second harmonic generation (SHG) coefficient $\beta_{ijk}(-2\omega;\omega,\omega)$,
the optical rectification (OR) coefficient $\beta_{ijk}(0;\omega,-\omega)$, the third
harmonic generation (THG) coefficient $\gamma_{ijkl}(-3\omega;\omega,\omega,\omega)$ and
the averaged degenerate four-wave mixing (DFWM) coefficient
$\bar{\gamma}_{ijkl}(-\omega;\omega,\omega,-\omega)$.~\cite{boyd2008nonlinear, takimoto2007real, Ding2013}

\begin{figure}
    \centering
    \includegraphics[width=0.95\textwidth]{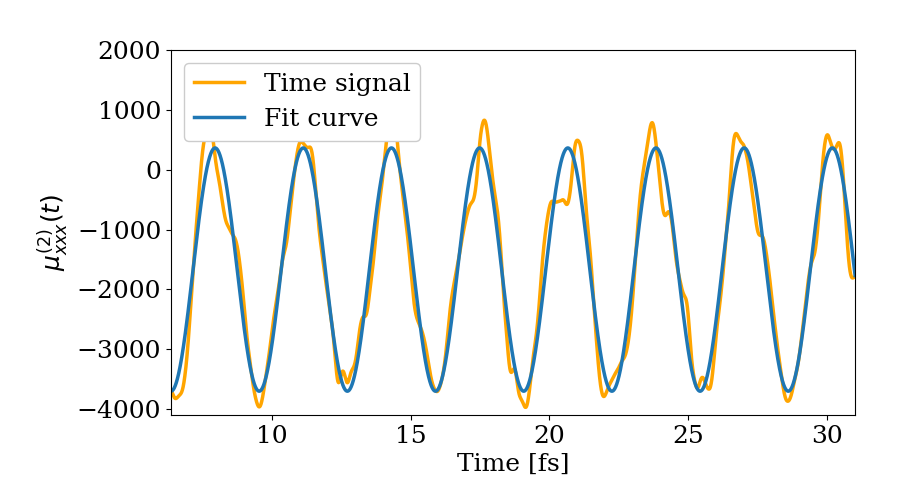}
    \caption{Non-linear optical properties from RT-TDDFT:
    fitting of second-order induced dipole moment contribution $\mu_{xxx}^{(2)}(t)$ of \ce{W(CO)5py}
    obtained as a numerical second derivative in each time step to analytical expression
    $\mu_{xxx}^{(2)}(t) = \tfrac{1}{4} \left[ \beta^\mathrm{SHG}_{xxx} \cos(2\omega t) + \beta^\mathrm{OR}_{xxx} \right]$
    from which elements of the hyperpolarizability tensor can be extracted. This figure is generated by the authors based on data underpinning Ref.~[\citenum{Konecny2016}].}
    \label{fig:NLOfitting}
\end{figure}

The workflow of the procedure for evaluating NLO properties from real-time simulations~\cite{Ding2013, Konecny2016} is as follows
\begin{enumerate}
    \item Starting from a converged ground-state SCF, perform several real-time simulations employing a cosine-shaped
    external field with different amplitudes of the field, for example
    $\mathcal{E}_1 = \mathcal{E}$, $\mathcal{E}_2 = 2\mathcal{E}$, $\mathcal{E}_3 = -\mathcal{E}$ and $\mathcal{E}_4
= -2\mathcal{E}$.
    Note that to improve the stability of time evolution and smoothness of extracted responses,
the cosine function is multiplied with a linear envelope $\omega t / (2\pi)$ in the first period.~\cite{Ding2013}
Different envelopes with improved performance has also been suggested.~\cite{Ofstad2023}
    \item Calculate $\mu^{(n)}$ as derivatives of induced dipole moment
    \begin{equation}
        \mu_{ij}^{(1)}(t)   = \left. \frac{\partial \mu_i(t)}{\partial \mathcal{E}_j} \right|_{\boldsymbol{\mathcal{E}}=0}, \quad
        \mu_{ijk}^{(2)}(t)  = \left. \frac{1}{2} \frac{\partial^2 \mu_i(t)}{\partial \mathcal{E}_j \partial \mathcal{E}_k} \right|_{\boldsymbol{\mathcal{E}}=0}, \quad
        \mu_{ijkl}^{(3)}(t) = \left. \frac{1}{6} \frac{\partial^2 \mu_i(t)}{\partial \mathcal{E}_j \partial \mathcal{E}_k \partial \mathcal{E}_l} \right|_{\boldsymbol{\mathcal{E}}=0},
    \end{equation}
    by means of numerical differentiation -- a finite field method in each time step.
    For example, the first- and second-order responses can be calculated from simulations emplying fields from step 1)
    with precision of the order $\mathcal{E}^4$ via
    \begin{subequations}
    \begin{flalign}
    \label{eq:finiteField1}
    \mu_{ij}^{(1)}(t)
    & = \frac{ 8  \left[ \mu_{i}(t,\mathcal{E}_j) - \mu_{i}(t,-\mathcal{E}_j) \right]  - \left[ \mu_{i}(t,2\mathcal{E}_j) - \mu_{i}(t,-2\mathcal{E}_j) \right] }{12\mathcal{E}_j},  \\
    \label{eq:finiteField2}
    \mu_{ijj}^{(2)}(t) & = \frac{ 16 \left[ \mu_{i}(t,\mathcal{E}_j) + \mu_{i}(t,-\mathcal{E}_j) \right]  - \left[ \mu_{i}(t,2\mathcal{E}_j) + \mu_{i}(t,-2\mathcal{E}_j) \right] }{24\mathcal{E}_j^2}.
    \end{flalign}
    \end{subequations}
    \item Fit the obtained $n$-th order induced dipole moment contributions to analytical expressions in Eqs.~\eqref{eq:NorderFD} to
    evaluate numerical values of the NLO properties.
\end{enumerate}
An example of such a real-time finite field procedure is depicted in Figure~\ref{fig:NLOfitting} for the second-order response $\mu_{xxx}^{(2)}(t)$
of \ce{W(CO)5py}, py = pyridine at the 1eX2C level of theory
(B3LYP functional~\cite{slater1951, vosko1980, Becke1988, Lee1988, Stephens1994}, Dyall’s uncontracted valence DZ basis set~\cite{dyall2010-5d} for W,
uncontracted aug-cc-pVDZ basis~\cite{Kendall1992} for the light elements).
The fitting was used to determine the second harmonic generation and optical rectification coefficients $\beta^\mathrm{SHG}_{xxx}$ and $\beta^\mathrm{OR}_{xxx}$,
respectively. The figure is based on data underpinning Ref.~[\citenum{Konecny2016}] where it was shown that the inclusion of relativistic effects contributed to about 35\% of the 
final value, highlighting the importance of a relativistic treatment in NLO applications where heavy metal-containing compounds are of interest due to the
favourable electronic properties of the metallic centre~\cite{Long1995, Bella2001}.

A special category of non-linear phenomena is high harmonic generation (HHG) during which photons from
a strong laser recombine into fewer photons of higher energy via an interaction with a material.
A HHG spectrum thus contains peaks corresponding to multiples of the frequency of the laser field.
HHG has practical importance both as a spectroscopy technique as well as a means of generating
coherent high-frequency radiation.
HHG presents a challenge for theoretical modelling due to the necessity of using strong fields
-- several orders of magnitude stronger than the applications discussed so far, thus requiring
a stable propagation, and due to the
requirements on basis sets that need to be able to describe electrons oscillating far from nuclei.
The first relativistic RT-TDDFT calculations of HHG were presented by De Santis et al. using
the PyBerthaRT program for \ce{Au2}, capturing harmonics up to the 13th order.~\cite{DeSantis2020pyberthart}

\subsection{Non-equilibrium spectroscopies}

So far we have discussed molecular properties where formulations in terms of perturbation theory
exist, and are usually the preferred mode of calculation. However, real-time methods are particularly well
suited for the simulation of experiments where the use of response theory would be too cumbersome.
Such is the case of non-equilibrium spectroscopies where more than one laser pulses are used to
drive the molecule. In the so-called pump--probe or transient absorption (TA) spectroscopies,
the first pulse (the pump) is used to excite the molecule into a non-equlibrium state while the
second pulse (the probe) then measures the response of the driven molecule. By varying the time
delay between the pump and the probe it is possible to follow quantum dynamics of electrons
in molecules in real time.
While a response theory-based description for pump--probe experiments exists in the form of
non-equlibrium response theory, 
the ability of real-time methods to tailor the pulse shape to match the experiments
and handle strong fields offers a distinct advantage over perturbative techniques.

In the case of transient absorption spectroscopies, two external pulses are used in the simulation,
the pump $\bm{\mathcal{E}}(t)$ and the probe $\bm{\mathcal{F}}(t)$ as introduced in the Fock matrix
in Eq.~\eqref{eq:4cFock}.
The pump first excites the molecule
to a non-stationary excited state. This perturbed state then evolves in time and its evolution is
probed by the second pulse applied after a time delay $\tau$.
As an example, let us consider a set-up with the pump pulse taking the form
\begin{equation}
	\bm{\mathcal{E}}(t)
	=
  \bm{n} \mathcal{E}(t)
  =
	\bm{n} \mathcal{E}
	\cos^2 \left(\pi \frac{t-t_0}{T} \right)
	\sin(\omega_0 t + \phi),
	\label{eq:pump}
\end{equation}
with amplitude $\mathcal{E}$, polarization direction $\vec{n}$, and the
pulse shape defined by the carrier frequency $\omega_0$ of a sine wave,
$\cos^2$ envelope, carrier--envelope phase (CEP) $\phi$, and time duration $T$.
The carrier frequency is usually tuned to an excitation energy of the molecule
which then becomes the primary excited state in the superposition state created by
the pump. However, even with relatively large amplitudes, the ground state
remains the most populated one.

For the probe, we use a broadband $\delta$-function pulse
\begin{equation}
	\bm{\mathcal{F}}(t)
	=
	\bm{m} \mathcal{F}(t)
	=
	\bm{m} \mathcal{F}_0 \delta(t-(T+\tau)),
	\label{eq:probe}
\end{equation}
applied at time $\tau$ after the pump pulse,
that similarly to the case of linear spectroscopies induces a time-dependent
dipole moment that can be processed to yield an absorption spectrum.
However, since the initial state now corresponds to the superposition state
instead of pure ground state, the spectrum contains imprints of quantum dynamics
of the non-equilibrium state.
The probe pulse can be applied while the pump is still active, overlapping regime,
or after the pump has been turned off, non-overlapping regime.

The final TA spectra are obtained within the RT-TDDFT framework from the differential induced dipole moment
\begin{equation}
  \label{eq:deltamu_TAS}
	\Delta \mu^{\mathrm{TAS}}_{uv}(t) 
  = 
  \Tr\left\{\mathbf{P}_u \left[\mathbf{D}_v^{\mathrm{pp}}(t) - \mathbf{D}_v^{\mathrm{p}}(t)\right]\right\} \equiv \mu^{\mathrm{ind,pp}}_{uv}(t) - \mu^{\mathrm{ind,p}}_{uv}(t),
  \qquad
  u,v \in \{x,y,z\},
\end{equation}
where $\mu^{\mathrm{ind}}_{uv}(t)$ denotes the expectation value of the dipole
operator. The computation of TA spectra involves performing two simulations
for recording the dipole moment at each time step; these simulations and their
quantities are denoted by p and pp subscripts, indicating that the real-time
propagation used pump-only pulse and pump together with the probe pulse,
respectively.

\begin{figure}[htb!]
	\centering
	\includegraphics[width=0.95\textwidth]{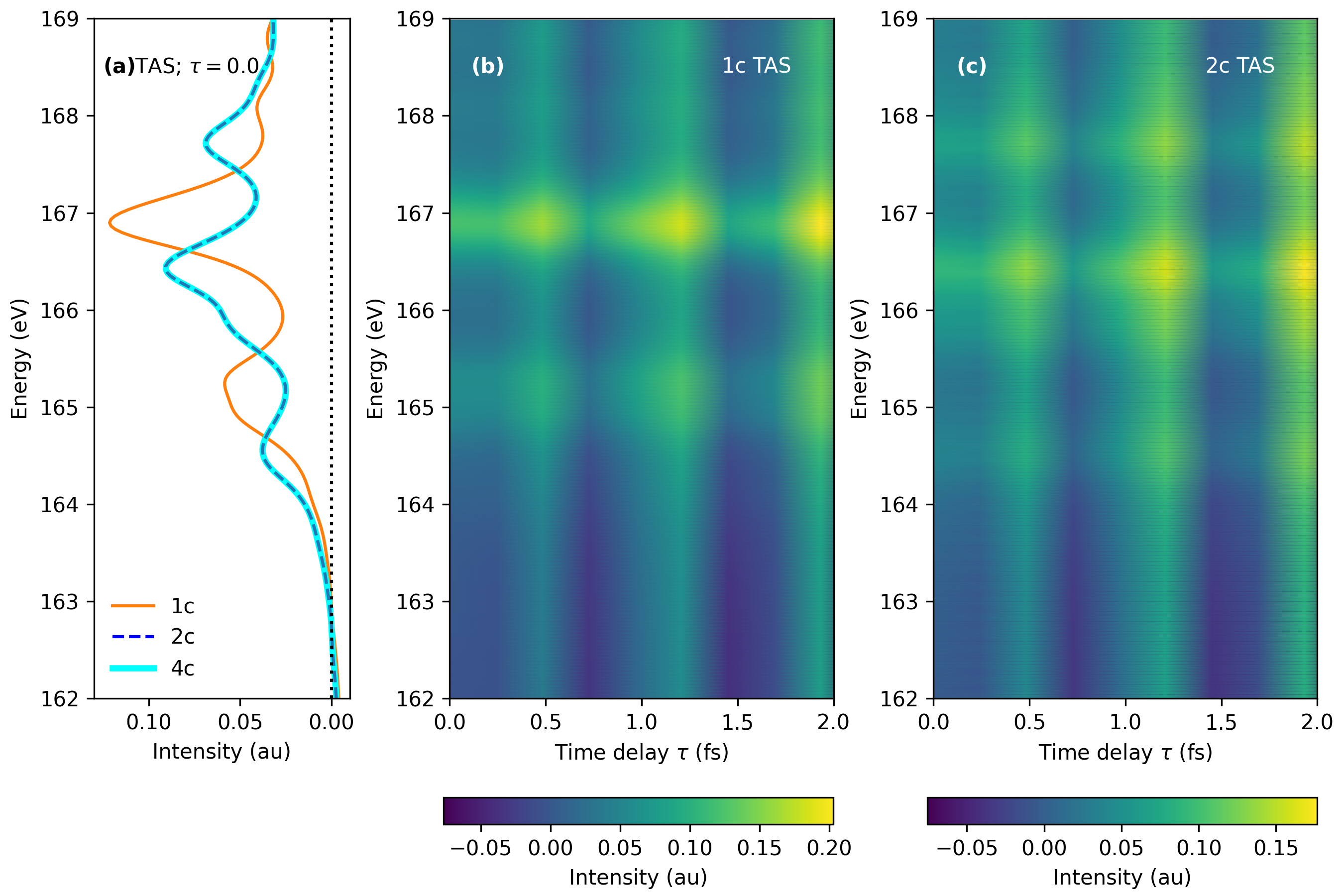}
	\caption{Transient absorption spectra of thiophene -- comparison of 1c non-relativistic (\emph{orange}),
        2c amfX2C (\emph{blue}) and 4c Dirac--Coulomb (\emph{cyan}) Hamiltonian-based RT-TDDFT.
		(a) TAS at $\tau=0$;
		(b) Variation in 1c TAS spectra with pump--probe delay $\tau$;
		(c) Variation in 2c TAS spectra with pump--probe delay $\tau$.
		All spectra are obtained with a damping factor $\Gamma = 0.01$ au.
        Figure reprinted from Ref.~[\citenum{moitra2023accurate}] under CC-BY-4.0 licence.
	}
	\label{fig:thiophene-hm}
\end{figure}

As an example, let us consider the TA spectrum of thiophene
(PBE0~\cite{slater1951, Perdew1996, Perdew1997, Adamo1999} functional modified to contain 40\% of Hartree--Fock exchange,
uncontracted aug-cc-pVXZ, X = T (S), D (C,H), basis~\cite{Dunning1989, Kendall1992, Woon1993})
depicted in Figure~\ref{fig:thiophene-hm}.
Here, the pump carrier frequency was set to correspond to the first excitation energy while the
X-ray absorption at sulfur \ce{L_{2,3}}-edges was investigated after the application of the
probe. The spin--orbit splitting of the sulfur 2p orbitals is visible in the spectrum which is
thus correctly described only by relativistic methods, in this case using the 4c Dirac--Coulomb
Hamiltonian (4c) and the amfX2C Hamiltonian (2c).
The pump--probe delay $\tau$ adds an extra degree of freedom to the TA spectrum which is then normally
plotted as a heat map where alternating low- and high-intensity signals can be observed, tracing
the dynamics of the superposition state as induced by the pump pulse.
Due to the increased computational cost of obtaining such a spectrum -- several spectra need to be
calculated from simulations with different $\tau$ in order to generate such a heat map -- efficient
2c relativistic methods are mandatory for these applications.

\section{Conclusion and perspectives}

Real-time methods are based on a direct integration of the quantum mechanical equation of motion.
In non-relativistic quantum chemistry, they gained popularity in previous decades due to their
ability to describe phenomena ranging from linear response properties to interaction with strong
laser fields and time-resolved spectroscopies -- areas at the forefront of experimental research.
In relativistic quantum chemistry, the pioneering theory development, computational implementation,
and first applications arrived later and the field has yet to catch up with the breadth of the scope
of applications of its non-relativistic counterparts.

In this chapter we summarized some of the advances of relativistic real-time methods in quantum chemistry
while restricting ourselves to mean-field methods and pure electron dynamics. It has been our ambition
that our introduction explains the fundamental principles of this methodology and inspires readers to join
in this rapidly developing and promising field.

 \bibliographystyle{elsarticle-num} 
 \bibliography{references}





\end{document}